\documentclass[10pt,conference]{IEEEtran}
\usepackage[utf8]{inputenc}
\usepackage{textcomp}
\IEEEoverridecommandlockouts

\usepackage{amsmath,amssymb,amsfonts}
\usepackage{graphicx}
\usepackage{xcolor}
\usepackage{algorithmic}
\usepackage{textcomp}
\usepackage{tabularx}
\usepackage{booktabs}
\usepackage{multicol, multirow}
\usepackage{xspace}
\usepackage{hyperref, url}
\usepackage{lipsum}
\usepackage{pgfplotstable}
\usepackage{pgfplots}
\usepgfplotslibrary{groupplots}
\pgfplotsset{compat=1.18}
\pgfplotsset{table/col sep=comma}
\usepackage{calc}
\usepackage{csvsimple}
\usepackage{pifont}
\usepackage{stfloats}
\usepackage[skins]{tcolorbox}
\usepackage{listings}
\usepackage{subfigure}
\usetikzlibrary{intersections}
\usepackage{venndiagram}
\usepackage{wrapfig}
\usetikzlibrary{calc}
\usepackage{amssymb}
\usepackage{makecell}

\def\totalExpressions{2,037\xspace}

\def\avgAccuracyOriginalImproved{86.2\%\xspace}

\def\percentImprovedClaudeOpusOS{45.5\%\xspace}

\def\percentImprovedClaudeOpusMS{44.6\%\xspace}

\def\percentImprovedGPTOSSOS{42.5\%\xspace}

\def\percentImprovedGPTOSSMS{42.3\%\xspace}

\def\percentFailedPhiOS{86.5\%\xspace}

\def\percentFailedPhiMS{83.4\%\xspace}

\def\totalExprsVTwoClaudeOpusOS{2,037\xspace}

\def\sOnePassVTwoClaudeOpusOS{1,684\xspace}
\def\percentSOnePassVTwoClaudeOpusOS{82.7\%\xspace}
\def\sOneRejVarsVTwoClaudeOpusOS{4\xspace}
\def\percentSOneRejVarsVTwoClaudeOpusOS{0.2\%\xspace}
\def\sOneRejCondsVTwoClaudeOpusOS{323\xspace}
\def\percentSOneRejCondsVTwoClaudeOpusOS{15.9\%\xspace}
\def\sOneRejBothVTwoClaudeOpusOS{10\xspace}
\def\percentSOneRejBothVTwoClaudeOpusOS{0.5\%\xspace}
\def\sOneRejOtherVTwoClaudeOpusOS{0\xspace}
\def\percentSOneRejOtherVTwoClaudeOpusOS{0\%\xspace}

\def\sOnePassVTwoClaudeOpusMS{1,705\xspace}
\def\percentSOnePassVTwoClaudeOpusMS{83.7\%\xspace}
\def\sOneRejVarsVTwoClaudeOpusMS{4\xspace}
\def\percentSOneRejVarsVTwoClaudeOpusMS{0.2\%\xspace}
\def\sOneRejCondsVTwoClaudeOpusMS{304\xspace}
\def\percentSOneRejCondsVTwoClaudeOpusMS{14.9\%\xspace}
\def\sOneRejBothVTwoClaudeOpusMS{11\xspace}
\def\percentSOneRejBothVTwoClaudeOpusMS{0.5\%\xspace}
\def\sOneRejOtherVTwoClaudeOpusMS{0\xspace}
\def\percentSOneRejOtherVTwoClaudeOpusMS{0\%\xspace}

\def\sOnePassVTwoGPTminiOS{1,765\xspace}
\def\percentSOnePassVTwoGPTminiOS{86.6\%\xspace}
\def\sOneRejVarsVTwoGPTminiOS{44\xspace}
\def\percentSOneRejVarsVTwoGPTminiOS{2.2\%\xspace}
\def\sOneRejCondsVTwoGPTminiOS{57\xspace}
\def\percentSOneRejCondsVTwoGPTminiOS{2.8\%\xspace}
\def\sOneRejBothVTwoGPTminiOS{83\xspace}
\def\percentSOneRejBothVTwoGPTminiOS{4.1\%\xspace}
\def\sOneRejOtherVTwoGPTminiOS{71\xspace}
\def\percentSOneRejOtherVTwoGPTminiOS{3.5\%\xspace}

\def\sOnePassVTwoGPTminiMS{1,800\xspace}
\def\percentSOnePassVTwoGPTminiMS{88.4\%\xspace}
\def\sOneRejVarsVTwoGPTminiMS{40\xspace}
\def\percentSOneRejVarsVTwoGPTminiMS{2\%\xspace}
\def\sOneRejCondsVTwoGPTminiMS{23\xspace}
\def\percentSOneRejCondsVTwoGPTminiMS{1.1\%\xspace}
\def\sOneRejBothVTwoGPTminiMS{88\xspace}
\def\percentSOneRejBothVTwoGPTminiMS{4.3\%\xspace}
\def\sOneRejOtherVTwoGPTminiMS{70\xspace}
\def\percentSOneRejOtherVTwoGPTminiMS{3.4\%\xspace}

\def\sOnePassVTwoGPTOSSOS{1,732\xspace}
\def\percentSOnePassVTwoGPTOSSOS{85\%\xspace}
\def\sOneRejVarsVTwoGPTOSSOS{6\xspace}
\def\percentSOneRejVarsVTwoGPTOSSOS{0.3\%\xspace}
\def\sOneRejCondsVTwoGPTOSSOS{273\xspace}
\def\percentSOneRejCondsVTwoGPTOSSOS{13.4\%\xspace}
\def\sOneRejBothVTwoGPTOSSOS{10\xspace}
\def\percentSOneRejBothVTwoGPTOSSOS{0.5\%\xspace}
\def\sOneRejOtherVTwoGPTOSSOS{1\xspace}
\def\percentSOneRejOtherVTwoGPTOSSOS{0\%\xspace}

\def\sOnePassVTwoGPTOSSMS{1,758\xspace}
\def\percentSOnePassVTwoGPTOSSMS{86.3\%\xspace}
\def\sOneRejVarsVTwoGPTOSSMS{8\xspace}
\def\percentSOneRejVarsVTwoGPTOSSMS{0.4\%\xspace}
\def\sOneRejCondsVTwoGPTOSSMS{245\xspace}
\def\percentSOneRejCondsVTwoGPTOSSMS{12\%\xspace}
\def\sOneRejBothVTwoGPTOSSMS{13\xspace}
\def\percentSOneRejBothVTwoGPTOSSMS{0.6\%\xspace}
\def\sOneRejOtherVTwoGPTOSSMS{6\xspace}
\def\percentSOneRejOtherVTwoGPTOSSMS{0.3\%\xspace}

\def\sOnePassVTwoPhiOS{1,681\xspace}
\def\percentSOnePassVTwoPhiOS{82.5\%\xspace}
\def\sOneRejVarsVTwoPhiOS{31\xspace}
\def\percentSOneRejVarsVTwoPhiOS{1.5\%\xspace}
\def\sOneRejCondsVTwoPhiOS{17\xspace}
\def\percentSOneRejCondsVTwoPhiOS{0.8\%\xspace}
\def\sOneRejBothVTwoPhiOS{46\xspace}
\def\percentSOneRejBothVTwoPhiOS{2.3\%\xspace}
\def\sOneRejOtherVTwoPhiOS{49\xspace}
\def\percentSOneRejOtherVTwoPhiOS{2.4\%\xspace}

\def\sOnePassVTwoPhiMS{1,680\xspace}
\def\percentSOnePassVTwoPhiMS{82.5\%\xspace}
\def\sOneRejVarsVTwoPhiMS{11\xspace}
\def\percentSOneRejVarsVTwoPhiMS{0.5\%\xspace}
\def\sOneRejCondsVTwoPhiMS{12\xspace}
\def\percentSOneRejCondsVTwoPhiMS{0.6\%\xspace}
\def\sOneRejBothVTwoPhiMS{38\xspace}
\def\percentSOneRejBothVTwoPhiMS{1.9\%\xspace}
\def\sOneRejOtherVTwoPhiMS{48\xspace}
\def\percentSOneRejOtherVTwoPhiMS{2.4\%\xspace}

\def\sTwoTotalPassedVTwoClaudeOpusOS{1,094\xspace}

\def\percentSTwoTotalPassedOfSOnePassVTwoClaudeOpusOS{65\%\xspace}
\def\sTwoEquivVTwoClaudeOpusOS{936\xspace}
\def\percentSTwoEquivVTwoClaudeOpusOS{45.9\%\xspace}
\def\percentSTwoEquivOfSOnePassVTwoClaudeOpusOS{55.6\%\xspace}
\def\sTwoDomainExpandVTwoClaudeOpusOS{158\xspace}

\def\percentSTwoDomainExpandOfSOnePassVTwoClaudeOpusOS{9.4\%\xspace}
\def\sTwoNotEquivVTwoClaudeOpusOS{574\xspace}

\def\percentSTwoNotEquivOfSOnePassVTwoClaudeOpusOS{34.1\%\xspace}
\def\sTwoDomainShrinkVTwoClaudeOpusOS{16\xspace}

\def\percentSTwoDomainShrinkOfSOnePassVTwoClaudeOpusOS{1\%\xspace}
\def\sTwoTotalPassedVTwoClaudeOpusMS{1,078\xspace}

\def\percentSTwoTotalPassedOfSOnePassVTwoClaudeOpusMS{63.2\%\xspace}
\def\sTwoEquivVTwoClaudeOpusMS{898\xspace}

\def\percentSTwoEquivOfSOnePassVTwoClaudeOpusMS{52.7\%\xspace}
\def\sTwoDomainExpandVTwoClaudeOpusMS{180\xspace}

\def\percentSTwoDomainExpandOfSOnePassVTwoClaudeOpusMS{10.6\%\xspace}
\def\sTwoNotEquivVTwoClaudeOpusMS{606\xspace}

\def\percentSTwoNotEquivOfSOnePassVTwoClaudeOpusMS{35.5\%\xspace}
\def\sTwoDomainShrinkVTwoClaudeOpusMS{21\xspace}

\def\percentSTwoDomainShrinkOfSOnePassVTwoClaudeOpusMS{1.2\%\xspace}
\def\sTwoTotalPassedVTwoGPTminiOS{464\xspace}

\def\percentSTwoTotalPassedOfSOnePassVTwoGPTminiOS{26.3\%\xspace}
\def\sTwoEquivVTwoGPTminiOS{431\xspace}

\def\percentSTwoEquivOfSOnePassVTwoGPTminiOS{24.4\%\xspace}
\def\sTwoDomainExpandVTwoGPTminiOS{33\xspace}

\def\percentSTwoDomainExpandOfSOnePassVTwoGPTminiOS{1.9\%\xspace}
\def\sTwoNotEquivVTwoGPTminiOS{869\xspace}

\def\percentSTwoNotEquivOfSOnePassVTwoGPTminiOS{49.2\%\xspace}
\def\sTwoDomainShrinkVTwoGPTminiOS{432\xspace}

\def\percentSTwoDomainShrinkOfSOnePassVTwoGPTminiOS{24.5\%\xspace}
\def\sTwoTotalPassedVTwoGPTminiMS{426\xspace}

\def\percentSTwoTotalPassedOfSOnePassVTwoGPTminiMS{23.7\%\xspace}
\def\sTwoEquivVTwoGPTminiMS{383\xspace}

\def\percentSTwoEquivOfSOnePassVTwoGPTminiMS{21.3\%\xspace}
\def\sTwoDomainExpandVTwoGPTminiMS{43\xspace}

\def\percentSTwoDomainExpandOfSOnePassVTwoGPTminiMS{2.4\%\xspace}
\def\sTwoNotEquivVTwoGPTminiMS{915\xspace}

\def\percentSTwoNotEquivOfSOnePassVTwoGPTminiMS{50.8\%\xspace}
\def\sTwoDomainShrinkVTwoGPTminiMS{459\xspace}

\def\percentSTwoDomainShrinkOfSOnePassVTwoGPTminiMS{25.5\%\xspace}
\def\sTwoTotalPassedVTwoGPTOSSOS{1,064\xspace}

\def\percentSTwoTotalPassedOfSOnePassVTwoGPTOSSOS{61.4\%\xspace}
\def\sTwoEquivVTwoGPTOSSOS{894\xspace}

\def\percentSTwoEquivOfSOnePassVTwoGPTOSSOS{51.6\%\xspace}
\def\sTwoDomainExpandVTwoGPTOSSOS{170\xspace}

\def\percentSTwoDomainExpandOfSOnePassVTwoGPTOSSOS{9.8\%\xspace}
\def\sTwoNotEquivVTwoGPTOSSOS{639\xspace}

\def\percentSTwoNotEquivOfSOnePassVTwoGPTOSSOS{36.9\%\xspace}
\def\sTwoDomainShrinkVTwoGPTOSSOS{29\xspace}

\def\percentSTwoDomainShrinkOfSOnePassVTwoGPTOSSOS{1.7\%\xspace}
\def\sTwoTotalPassedVTwoGPTOSSMS{1,060\xspace}

\def\percentSTwoTotalPassedOfSOnePassVTwoGPTOSSMS{60.3\%\xspace}
\def\sTwoEquivVTwoGPTOSSMS{860\xspace}

\def\percentSTwoEquivOfSOnePassVTwoGPTOSSMS{48.9\%\xspace}
\def\sTwoDomainExpandVTwoGPTOSSMS{200\xspace}

\def\percentSTwoDomainExpandOfSOnePassVTwoGPTOSSMS{11.4\%\xspace}
\def\sTwoNotEquivVTwoGPTOSSMS{664\xspace}

\def\percentSTwoNotEquivOfSOnePassVTwoGPTOSSMS{37.8\%\xspace}
\def\sTwoDomainShrinkVTwoGPTOSSMS{34\xspace}

\def\percentSTwoDomainShrinkOfSOnePassVTwoGPTOSSMS{1.9\%\xspace}
\def\sTwoTotalPassedVTwoPhiOS{291\xspace}

\def\percentSTwoTotalPassedOfSOnePassVTwoPhiOS{17.3\%\xspace}
\def\sTwoEquivVTwoPhiOS{264\xspace}
\def\percentSTwoEquivVTwoPhiOS{13\%\xspace}
\def\percentSTwoEquivOfSOnePassVTwoPhiOS{15.7\%\xspace}
\def\sTwoDomainExpandVTwoPhiOS{27\xspace}

\def\percentSTwoDomainExpandOfSOnePassVTwoPhiOS{1.6\%\xspace}
\def\sTwoNotEquivVTwoPhiOS{941\xspace}

\def\percentSTwoNotEquivOfSOnePassVTwoPhiOS{56\%\xspace}
\def\sTwoDomainShrinkVTwoPhiOS{449\xspace}

\def\percentSTwoDomainShrinkOfSOnePassVTwoPhiOS{26.7\%\xspace}
\def\sTwoTotalPassedVTwoPhiMS{350\xspace}

\def\percentSTwoTotalPassedOfSOnePassVTwoPhiMS{20.8\%\xspace}
\def\sTwoEquivVTwoPhiMS{315\xspace}

\def\percentSTwoEquivOfSOnePassVTwoPhiMS{18.8\%\xspace}
\def\sTwoDomainExpandVTwoPhiMS{35\xspace}

\def\percentSTwoDomainExpandOfSOnePassVTwoPhiMS{2.1\%\xspace}
\def\sTwoNotEquivVTwoPhiMS{968\xspace}

\def\percentSTwoNotEquivOfSOnePassVTwoPhiMS{57.6\%\xspace}
\def\sTwoDomainShrinkVTwoPhiMS{362\xspace}

\def\percentSTwoDomainShrinkOfSOnePassVTwoPhiMS{21.5\%\xspace}

\def\percentSTwoSympyOfExprsVTwoClaudeOpusOS{6.5\%\xspace}

\def\percentSTwoSympyOfExprsVTwoPhiOS{2.1\%\xspace}

\def\avgAccuracyIncreaseClaudeOpusMS{10.2\%\xspace}
\def\avgAccuracyIncreaseGPTminiOS{5.4\%\xspace}
\def\avgAccuracyIncreaseGPTminiMS{6.6\%\xspace}

\def\avgAccuracyIncreaseHerbie{11.6\%\xspace}
\def\avgAccuracyIncreasePhiOS{7\%\xspace}
\def\avgAccuracyIncreasePhiMS{8.3\%\xspace}

\def\avgAccuracyIncreaseSixVarLLM{7\%\xspace}

\def\avgAccuracyIncreaseSevenVarHerbie{12.3\%\xspace}

\def\avgAccuracyIncreaseTenVarLLM{10.3\%\xspace}

\def\avgAccuracyIncreaseSixteenVarHerbie{11.5\%\xspace}

\def\avgAccuracyIncreaseFourVarGPTOSSMS{13.5\%\xspace}

\def\avgAccuracyIncreaseSevenVarHerbie{17.8\%\xspace}

\def\avgAccuracyIncreaseSixteenVarHerbie{8.5\%\xspace}

\def\maxConditionalLevel{6\xspace}

\def\avgAccuracyIncreaseCondLLM{7.6\%\xspace}
\def\avgAccuracyIncreaseCondHerbie{11.6\%\xspace}

\def\avgAccuracyIncreaseUncondLLM{9.2\%\xspace}

\def\avgAccuracyOriginalCond{86\%\xspace}
\def\avgAccuracyOriginalNonCond{86.5\%\xspace}

\def\avgAccuracyIncreaseCondHerbie{11.6\%\xspace}
\def\avgAccuracyIncreaseNonCondHerbie{11.5\%\xspace}

\def\maxOpsOverall{2474\xspace}
\def\avgOpsOverall{76.4\xspace}
\def\minOpsOneVar{3\xspace}
\def\maxOpsOneVar{67\xspace}
\def\avgOpsOneVar{20\xspace}
\def\minOpsTwoVar{2\xspace}
\def\maxOpsTwoVar{85\xspace}
\def\avgOpsTwoVar{17.9\xspace}
\def\minOpsThreeVar{3\xspace}
\def\maxOpsThreeVar{128\xspace}
\def\avgOpsThreeVar{26.3\xspace}
\def\minOpsFourVar{4\xspace}
\def\maxOpsFourVar{116\xspace}
\def\avgOpsFourVar{29\xspace}
\def\minOpsFiveVar{4\xspace}
\def\maxOpsFiveVar{168\xspace}
\def\avgOpsFiveVar{39\xspace}
\def\minOpsSixVar{3\xspace}
\def\maxOpsSixVar{187\xspace}
\def\avgOpsSixVar{46\xspace}
\def\minOpsSevenVar{8\xspace}
\def\maxOpsSevenVar{523\xspace}
\def\avgOpsSevenVar{93.6\xspace}
\def\minOpsEightVar{12\xspace}
\def\maxOpsEightVar{76\xspace}
\def\avgOpsEightVar{32.6\xspace}
\def\minOpsNineVar{13\xspace}
\def\maxOpsNineVar{451\xspace}
\def\avgOpsNineVar{101.7\xspace}
\def\minOpsTenVar{13\xspace}
\def\maxOpsTenVar{89\xspace}
\def\avgOpsTenVar{40.4\xspace}
\def\minOpsSixteenVar{94\xspace}
\def\maxOpsSixteenVar{2474\xspace}
\def\avgOpsSixteenVar{487.9\xspace}

\def\totalHighPrecisionOriginal{277\xspace}
\def\percentHighPrecisionOriginal{13.6\%\xspace}

\def\percentImprovedHighPrecisionClaudeOpusOS{18.4\%\xspace}

\def\percentImprovedHighPrecisionClaudeOpusMS{20.9\%\xspace}

\def\percentImprovedHighPrecisionGPTminiOS{11.9\%\xspace}

\def\percentImprovedHighPrecisionGPTminiMS{11.6\%\xspace}

\def\percentImprovedHighPrecisionGPTOSSOS{22\%\xspace}

\def\percentImprovedHighPrecisionGPTOSSMS{21.3\%\xspace}

\def\percentImprovedHighPrecisionHerbie{88.1\%\xspace}

\def\percentImprovedHighPrecisionPhiOS{8.3\%\xspace}

\def\percentImprovedHighPrecisionPhiMS{7.9\%\xspace}

\def\avgAccuracyIncreaseFullLLM{9.2\%\xspace}
\def\avgAccuracyIncreaseFullHerbie{11.6\%\xspace}
\def\avgAccuracyIncreaseMixedLLM{6.2\%\xspace}
\def\avgAccuracyIncreaseMixedHerbie{8.4\%\xspace}
\def\avgAccuracyIncreaseUnaryLLM{11.7\%\xspace}
\def\avgAccuracyIncreaseUnaryHerbie{14.9\%\xspace}

\def\avgAccuracyIncreaseFullHerbie{11.6\%\xspace}
\def\avgAccuracyIncreaseMixedHerbie{8.4\%\xspace}
\def\avgAccuracyIncreaseUnaryHerbie{14.9\%\xspace}

\def\totalLLMBetterThanHerbie{489\xspace}
\def\percentLLMBetterThanHerbie{24\%\xspace}
\def\totalHerbieNotImproved{129\xspace}
\def\percentHerbieNotImproved{6.3\%\xspace}
\def\totalLLMImprovedWhenHerbieNot{79\xspace}
\def\percentLLMImprovedWhenHerbieNot{61.2\%\xspace}
\def\percentLLMImprovedWhenHerbieNotOverall{3.9\%\xspace}

\def\percentHerbieBetterThanLLMOfComparable{42.7\%\xspace}

\def\vennTotal{2,037\xspace}

\def\vennLLMImprovesPct{67.2\%\xspace}

\def\vennHerbieImprovesPct{93.7\%\xspace}
\def\vennBoth{1,289\xspace}

\def\vennLLMOnly{79\xspace}

\def\vennHerbieOnly{619\xspace}

\def\taskOnePromptCountFull{380,000\xspace}
\def\taskOneTokentCount{300 million\xspace}

\def\taskTwoPromptCountFullNum{88,900\xspace}
\def\taskTwoTokentCount{2.7 billion\xspace}
\def\totalPromptCount{469,000\xspace}
\def\totalTokenCount{3 billion\xspace}
\def\avgAccuracyPassedSemantic{92.8\%\xspace}
\def\avgAccuracyFailedSemantic{72.3\%\xspace}

\def\BibTeX{{\rm B\kern-.05em{\sc i\kern-.025em b}\kern-.08em
    T\kern-.1667em\lower.7ex\hbox{E}\kern-.125emX}}
\begin{document}

\newtcolorbox{rqbox}{
  enhanced,
  width=\columnwidth,
  drop fuzzy shadow southwest,
  colframe=red!50!black,
  colback=yellow!10,
  left=0mm,
  right=0mm,
  top=0mm,
  bottom=0mm
}

\definecolor{c1}{HTML}{534AB7}
\definecolor{c2}{HTML}{AFA9EC}
\definecolor{c3}{HTML}{0F6E56}
\definecolor{c4}{HTML}{5DCAA5}
\definecolor{c5}{HTML}{993C1D}
\definecolor{c6}{HTML}{F0997B}
\definecolor{c7}{HTML}{185FA5}
\definecolor{c8}{HTML}{85B7EB}
\definecolor{c9}{HTML}{888780}
\definecolor{c10}{HTML}{EF9F27}
\definecolor{c11}{HTML}{D4537E}

\definecolor{codegreen}{rgb}{0,0.6,0}
\definecolor{codegray}{rgb}{0.5,0.5,0.5}
\definecolor{codepurple}{rgb}{0.58,0,0.82}
\definecolor{backcolour}{rgb}{0.95,0.95,0.92}

\lstdefinestyle{mystyle}{
    backgroundcolor=\color{backcolour},   
    commentstyle=\color{codegreen},
    keywordstyle=\color{magenta},
    numberstyle=\tiny\color{codegray},
    stringstyle=\color{codepurple},
    basicstyle=\ttfamily\small,
    breakatwhitespace=false,         
    breaklines=true,                 
    captionpos=b,                    
    keepspaces=true,                 
    numbers=left,                    
    numbersep=5pt,                  
    showspaces=false,                
    showstringspaces=false,
    showtabs=false,                  
    tabsize=2
}

\lstset{style=mystyle}

\newcommand{\gulzar}[1]{\textcolor{red}{G: #1}}
\newcommand{\tien}[1]{\textcolor{blue}{T: #1}}
\newcommand{\krish}[1]{\textcolor{cyan}{K: #1}}
\newcommand{\todo}[1]{\textcolor{green}{TODO: #1}}


\title{Assessing Large Language Models for Stabilizing Numerical Expressions in Scientific Software}

\author{Anonymous Authors}

\author{\IEEEauthorblockN{Tien Nguyen}
\IEEEauthorblockA{\textit{Department of Computer Science} \\
\textit{Virginia Tech}\\
Blacksburg, USA \\
tiennguyen@vt.edu}
\and
\IEEEauthorblockN{Kirshanthan Sundararajah}
\IEEEauthorblockA{\textit{Department of Computer Science} \\
\textit{Virginia Tech}\\
Blacksburg, USA \\
kirshanthans@vt.edu}
\and
\IEEEauthorblockN{Muhammad Ali Gulzar}
\IEEEauthorblockA{\textit{Department of Computer Science} \\
\textit{Virginia Tech}\\
Blacksburg, USA \\
gulzar@cs.vt.edu}
}

\maketitle

\begin{abstract}

    Scientific software relies on high-precision computation, yet finite floating-point representations can introduce precision errors that propagate in safety-critical domains. Despite the growing use of large language models (LLMs) in scientific applications, their reliability in handling floating-point numerical stability has not been systematically evaluated. This paper evaluates LLMs’ reasoning on high-precision numerical computation through two numerical stabilization tasks: (1) detecting instability in numerical expressions by generating error-inducing inputs ({\em detection}), and (2) rewriting expressions to improve numerical stability ({\em stabilization}). Building on popular numerical benchmarks, we assess 4 state-of-the-art LLMs on \totalExpressions numerical structures, including nested conditionals, high-precision literals, and multi-variable arithmetic, across \totalPromptCount tasks.
     
    Our results show that LLMs complement state-of-the-art traditional approaches in detecting and stabilizing numerically unstable computations. More notably, LLMs outperform baseline methods precisely where the latter fail, stabilizing \percentLLMImprovedWhenHerbieNot of the expressions that the baseline fails to improve. More broadly, however, traditional stabilization and detection baselines outperform LLMs. For instance, Herbie stabilizes \vennHerbieImprovesPct of the expressions compared to \vennLLMImprovesPct by LLMs, and on expressions that both stabilize, Herbie achieves higher accuracy in \percentHerbieBetterThanLLMOfComparable of the cases. LLMs struggle with control flow and high-precision literals, consistently removing such structures rather than reasoning about their numerical implications, while performing substantially better on purely symbolic expressions. Even when LLMs preserve structure, their numerical reasoning falters, yielding semantically inequivalent expressions in over 46\% of cases. These findings suggest LLMs are effective at stabilizing expressions that classical techniques cannot, yet struggle when high-precision magnitudes and control-flow semantics demand precise reasoning, since such concrete patterns are rarely seen during training.

\end{abstract}


\section{INTRODUCTION}
\label{s1:intro}

    Numerical faults are common in scientific software \cite{Davis2025, Franco2017}. Due to the finite nature of floating-point representation in computing, rounding errors, cancellation \cite{GeeksforGeeks2025}, overflow, underflow, and numerical instability \cite{numericalstability} cause results to deviate from the mathematically exact values. For example, in finite-precision arithmetic, computing $e^{1000}$ may overflow in standard floating-point, even though the mathematical result is finite. These issues can accumulate and lead to software faults, as evidenced by recent CVEs \cite{Pycytpes, SoX, Poppler} involving mishandled floating-point arithmetic expressions. Addressing this challenge typically requires reasoning over large input spaces to ensure no configuration triggers undefined behavior. 

    Take the quadratic formula $x = \frac{-b \pm \sqrt{b^2 - 4ac}}{2a}$ as a simple example. Despite being mathematically correct, its direct floating-point implementation can be numerically unstable, particularly when $\sqrt{b^2 - 4ac}$ becomes close to $b$ when $b^2 \gg 4ac$. In this situation, the computation involves subtracting nearly equal large numbers, causing catastrophic cancellation and loss of significant bits. For instance, with $a = c = 1$, when $b$ exceeds $10^5$, one root is computed as $x_1 = -0.00010000000111176632$, while the more stable formulation yields $x_1 = -0.00010000000100000001$. This discrepancy arises because the floating-point representation cannot accurately capture the small difference between the nearly equal terms, producing spurious digits due to rounding errors. In practice, such cases can also lead to finite-precision-based attacks \cite{Haney2022}, where inputs are deliberately structured to amplify rounding errors, cancellation, or instability under floating-point arithmetic to reveal failure modes.
    

    To mitigate these issues, prior work has explored automated detection and correction techniques, including precision tuning and algebraic rewriting. Tools such as Herbie \cite{herbie}, Daisy \cite{daisy}, and FPGen \cite{fpgen}  transform or analyze numerical expressions and programs to reduce rounding errors and improve stability. These approaches rely on domain-specific algorithms and handcrafted search strategies, forming a strong baseline for evaluating numerical correctness and stability.

    
   Large language models (LLMs) have demonstrated strong performance in symbolic reasoning and structured problem solving \cite{Pan2023, Kojima2023}. However, their reliability in scientific computing, particularly numerical reasoning over floating-point expressions, remains largely unexplored. Existing evidence suggests that performance on numerical tasks is sensitive to task complexity \cite{Dave2024}, and that incorrect outputs in testing or repair workflows can have serious consequences \cite{Diehl2025}. 
   \textit{We hypothesize that while LLMs possess symbolic reasoning capabilities, their performance may lack rigor in accurately assessing the magnitude, range, and sensitivity of numerical errors.} This hypothesis stems from how LLMs are designed. They extrapolate learned statistical patterns over symbolic expressions, enabling them to associate symbolic structures with known instability patterns. Training on exact, high-precision numerical values is largely absent, leaving LLMs unable to characterize the magnitude of the error.
    



    \noindent\textbf{\em Contributions.} We conduct a systematic evaluation of LLMs' ability to reason about numerical computations, focusing on the detection and mitigation of instability. Numerical stability can be viewed as a two-part problem, analogous to fault detection and repair: first, identifying inputs that trigger high floating-point error ({\em detection}), and second, generating transformations that reduce such errors ({\em stabilization}). We design two complementary experimental tasks that target different aspects of stability analysis. First, we introduce an error-inducing input generation task, which evaluates whether models can detect inputs that maximize floating-point error, using a state-of-the-art benchmark from prior work \cite{fpgen}. 
    Our analysis shows that numerical issues arise primarily from expression-level computations rather than the surrounding software. Motivated by this, the second task focuses on stabilizing floating-point expressions, where models generate transformations to improve numerical stability.
   
    Realizing these evaluations requires addressing four challenges. First, existing expression benchmarks and their stable variants may appear in LLM training data, causing data contamination that gives LLMs an unfair advantage. Second, these benchmarks offer limited control over expression complexity, hindering systematic evaluation. Third, they contain relatively few expressions, limiting generalizable conclusions. Fourth, LLMs may not preserve semantics during stabilization. To address these challenges, we construct a new dataset by generating composed expressions from existing source benchmarks \cite{herbie}. This method produces many previously unseen expressions with controllable complexity across arithmetic operations, symbolic variables, and control flows. We validate LLM-based stabilization using both structural checks (e.g., number of variables and control flows) and established, standardized numerical checks (e.g., domain and accuracy)~\cite{herbie, fpbench}.

       \noindent\textbf{\em Evaluation Setting.}  We evaluate four
       state-of-the-art LLMs, spanning open-source, commercial, and frontier models, in two phases: instability detection and stabilization. For detection, we use 21 source functions from prior work \cite{fpgen} using FPGen as the baseline, since it detects floating-point precision errors in mathematical libraries.
      For stabilization, we apply three composition techniques to expressions from the prior benchmark~\cite{herbie} and sample uniformly across them, varying the number of variables and conditional structures, yielding \totalExpressions expressions in total. In total, our experiments comprise approximately \totalPromptCount prompts across both tasks and \totalTokenCount tokens. No prior work in numerical stability has experimented at this scale.
    
     \noindent\textbf{\em Study Results.} We observe that LLMs are not replacements for classical baselines such as FPGen \cite{fpgen} and Herbie \cite{herbie} in detecting and stabilizing numerical instability, but they offer complementary benefits. 
    In the detection task, LLMs match or exceed approximately half of the benchmarks in generating inputs that maximize the relative error between high-precision and float64 computation. Notably, some models, such as Phi4 and GPT-OSS, can exploit the numerical structure (e.g., QR decomposition) to induce extremely large relative errors. Surprisingly, their performance deteriorates on functions with simpler computations, such as summations, vector dot products, and convolutions.

    In the stabilization task, the baseline~\cite{herbie} consistently outperforms the LLMs. LLMs achieve numerical stabilization for \vennLLMImprovesPct of expressions in the dataset, with accuracy improvements ranging from \avgAccuracyIncreaseGPTminiOS to \avgAccuracyIncreaseClaudeOpusMS, compared to the baseline, which achieves numerical stabilization on \vennHerbieImprovesPct of the expressions in the dataset, and with an average accuracy improvement of \avgAccuracyIncreaseHerbie. Thus, baselines not only stabilize more expressions but also offer better accuracy improvement (i.e., \avgAccuracyIncreaseHerbie by Herbie vs. \avgAccuracyIncreaseClaudeOpusMS by Opus 4.8).
    However, the expressions stabilized by the LLM do not fully overlap with those stabilized by the baseline. In \percentHerbieNotImproved (\totalHerbieNotImproved) of expressions where the baseline fails to improve accuracy, the LLM successfully stabilizes \totalLLMImprovedWhenHerbieNot (\percentLLMImprovedWhenHerbieNot) of them. These are typically large, multi-variable expressions whose fix requires a single named identity spanning several library functions (e.g., $\tanh$, $\mathrm{atanh}$, $\cos 2x$), which the baseline's arithmetic-level rewrite rules cannot express. Similarly, LLM achieves greater stability than the baseline for \percentLLMBetterThanHerbie (\totalLLMBetterThanHerbie) of all expressions. In these cases, the baseline finds a valid but partial fix, while the LLM applies one transformation across the entire expression and removes a larger portion of the error at once.

    In over 46\% of the cases, LLMs did not preserve the semantics, and their performance degrades as structural complexity increases, including sensitivity to branching, composition depth, and precision interactions. Notable mistakes include dropping variables, literals, and conditions, as well as shrinking the domain on which the expressions operate. Among such expressions, up to $17.5\%$ had dropped variable and/or conditions, and an even larger portion ($46.3-85.7\%$) had large accuracy gaps, indicating that LLMs lack high-precision numerical reasoning ability. On expressions containing high-precision literals, LLM accuracy improvement ranges from \percentImprovedHighPrecisionPhiMS to \percentImprovedHighPrecisionGPTOSSOS, compared to \percentImprovedHighPrecisionHerbie for the baseline, indicating that LLMs struggle to reason about high-precision literals. This is consistent with our hypothesis that LLMs lack the rigor to accurately assess the magnitude, range, and sensitivity of numerical errors. Exact high-precision literals are rarely encountered as concrete values during training, whereas LLMs recognize control-flow syntax but fail to reason about their implications for numerical stability. We also observe that few-shot prompting generally improves performance over zero-shot across all examined structural categories, and that larger, frontier-level (e.g., Opus 4.8) models outperform smaller models. This finding suggests that LLMs lack stable reasoning about numerical behavior across computations and should therefore be treated as stochastic approximators that complement, rather than replace, domain-specific tools for high-precision computation.

\section{MOTIVATION}
\label{s3:motivation}

    This section presents two case studies illustrating scenarios in which LLMs both outperform and underperform baseline~\cite{herbie} in rewriting expressions into stable forms.


    \subsection{Case 1: Successful Numerical Stabilization by LLMs}


Consider the expression $(1-\cos u)\left(\frac{1}{u}\right)^2$, where $u=\frac{\ln(1-\frac{\ln(1-x)}{\ln(1+x)})}{\ln(1+\frac{\ln(1-x)}{\ln(1+x)})}$. This is a single-variable, non-conditional expression generated using the unary composition strategy (explained in Section~\ref{s5.2:composition}), with $x$ defined over the full-range domain.\footnote{The full-range domain is defined as $[-1.79\times10^{308},\, 1.79\times10^{308}]$ and is discussed in Section~\ref{s4:methodology}.} Under the baseline error metric~\cite{herbie, STOKE}, the original expression exhibits an average error of 4.5 bits, roughly the precision lost to floating-point rounding when comparing 256-bit and 64-bit evaluations across 256 sampling points, corresponding to an average accuracy of 91.5\% and indicating numerical instability. Formal definitions of these metrics are provided in Section~\ref{s4:methodology}.

The instability originates from the term $1-\cos u$. As $u \to 0$, $\cos u \to 1$, so the subtraction $1-\cos u$ cancels two nearly equal operands and suffers catastrophic cancellation. Claude Opus 4.8 rewrites this term using its mathematically equivalent form $2\sin^2(\frac{u}{2})$, which is free of such cancellation because it evaluates a small quantity directly rather than as the difference of two near-unity values. The model thus produces a more stable form of the original expression while preserving its exact value, reducing the error from 4.5 to 2.6 bits and improving accuracy from 91.5\% to 95.1\%.

The baseline~\cite{herbie}, by contrast, retains the structure of the original expression but replaces the inner arguments with truncated Taylor-series polynomial approximations, yielding $u =\frac{\ln(2 + x + \frac{1}{2}x^2 + \frac{5}{12}x^3)}{\ln(x(-1 -\frac{1}{2}x - \frac{5}{12}x^2 - \frac{7}{24}x^3))}$. This rewrite is not value-preserving. It introduces approximation error from the polynomial substitution, and it still retains the cancellation-prone $1-\cos u$ term. As a result, it incurs roughly 15.3 bits of error and only 71.2\% accuracy, which is worse than the original.

        \subsection{Case 2: Failure in Numerical Stabilization by LLMs}
    Case~1 showed that LLMs excel when stabilization reduces to applying a well-known, named identity. 
    Stabilization, however, does not always reduce to such a recognizable rewrite. When the fix instead requires reasoning over the specific structure of an expression---its branch thresholds, the magnitude of high-precision literals, and where significance is actually lost---the baseline~\cite{herbie} tends to dominate. For example, consider a two-variable conditional expression defined as follows.
    
     {
     \footnotesize
     
     $
     E =
        \begin{cases}
        x^{10} - \dfrac{y^3}{2x^{10}} & \text{if } x^{10} < -1.5097698010473000 \times 10^{29} \\
        \sqrt{x^{20} + y^3} & \text{if } x^{10} < 5.582399551122500 \times 10^{29} \\
        x^{10} + \dfrac{y^3}{2x^{10}} & \text{otherwise}
        \end{cases}
        $
     }
     
    This expression combines multiple conditional branches with extreme exponents and high-precision literals, a structure that admits no single textbook identity. Terms like $x^{10} \pm \frac{y^3}{2x^{10}}$ risk loss of significance when $x^{10} \gg \frac{y^3}{2x^{10}}$, $\sqrt{x^{20} + y^3}$ can overflow, and branching on $x^{10}$ amplifies sensitivity to rounding errors near the thresholds. Stabilizing it requires \emph{quantitative} reasoning about operand magnitudes rather than recognizing a known form. Herbie performs exactly this analysis, introducing additional case splits and algebraic transformations targeted at the regions where error concentrates. For example, it rewrites the term $x^{10} + \frac{y^3}{2x^{10}} \rightarrow\operatorname{fma}\!\big(\tfrac{1}{2} y^2 x^{-10},\, y,\, x^{10}\big)$, replacing a division and an addition with a fused multiply-add that preserves precision when a small term is added to a large one, improving accuracy by 21\%.
    
   Phi4, by contrast, attempts to recognize and substitute familiar algebraic forms. It rephrases the expression terms and conditions to look simpler without modeling its numerical behavior over the input domain. Lacking a named identity to apply, the model does not reason about the role of the high-precision literals, the relative magnitudes that drive cancellation, or the need to introduce precision-preserving primitives such as $\operatorname{fma}$ or new case splits to avoid overflow and underflow in extreme ranges. The resulting expression is 

     {
     \footnotesize
     $
           E =
        \begin{cases}
        |x|^{10} - \dfrac{|y^3|}{2|x|^{10}} & \text{if } x^{10} < -1.5097698010472593 \times 10^{80} \\
        \sqrt{|x|^{20} + |y|^{3}} & \text{if } x^{10} < 5.582399551122541 \times 10^{79} \\
        |x|^{10} + \dfrac{|y^3|}{2|x|^{10}} & \text{otherwise}
        \end{cases}
        $
    }

    \noindent with accuracy of 78.2\%, slightly worse than the original.

\section{METHODOLOGY}
\label{s4:methodology}

    To systematically evaluate the numerical stability capabilities of LLMs, we formulate the following research questions. 
    \begin{itemize}
        \item \textbf{RQ1:} To what extent can LLMs generate inputs that expose numerical errors in programs?
        \item \textbf{RQ2:} How accurately can we identify semantically divergent rewrites by LLMs?
        \item \textbf{RQ3:} How effectively can LLMs rewrite numerically unstable expressions to improve numerical accuracy?
        \item \textbf{RQ4:} How does expression structural complexity affect LLM performance?
    \end{itemize}

    \subsection{Floating-Point Instability Detection}

        \begin{figure}
    \centering
    \includegraphics[width=\columnwidth]{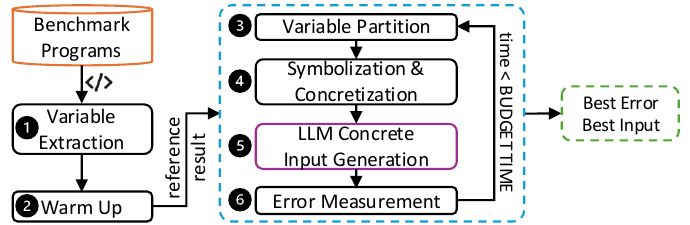}
    \vspace{-18pt}
    \caption{Instability detection workflow}
    \label{fig:Fig1}
    \vspace{-3ex}
\end{figure}    

        We first instruct LLMs to generate inputs that maximize floating-point errors, such as cancellation, overflow, and underflow, by increasing the relative error between high-precision and standard 64-bit executions. This setup directly measures whether LLMs can exploit common sources of instability. Figure \ref{fig:Fig1} illustrates the overall workflow.
        
        \subsubsection{Benchmark Acquisition} We use FPGen~\cite{fpgen} as the baseline, a state-of-the-art tool that  discovers numerical instability in popular mathematical libraries. The accompanying benchmark provides 21 programs from mathematical libraries (GSL, Meschach, and summation routines). As LLMs may have seen the baseline benchmarks' best error-inducing inputs during training, we convert them from C to Python for evaluation. The benchmark results are also reproducible from the functional FPGen software artifact. 

        \subsubsection{Experiment Design}
        For each program, we first extract all input variables (Step \ding{202}), treating each input as a symbolic variable. Following \cite{fpgen}'s warm-up phase, we randomly generate concrete input values to find the best inputs and the best error within a fixed time (Step \ding{203}).  
        Steps \ding{204}–\ding{207} replace symbolic execution with LLM-guided input generation. Similar to FPGen, we randomly partition the input variables into symbolic and concrete sets (Step \ding{204}). The concrete variables are fixed using values from the current best result (Step \ding{205}). 
        In Step \ding{206}, we prompt LLMs to generate candidate inputs for the symbolic variables in a feedback-guided prompting strategy, enabling LLMs to refine previously generated candidates and explore beyond known high-error regions. 
        Valid inputs are then evaluated on both high-precision and standard floating-point executions to compute relative error (Step \ding{207}). We use 256-bit precision for high-precision evaluation to obtain more accurate reference values and reduce numerical noise compared to FPGen’s 128-bit setting. The best observed error and corresponding inputs guide subsequent iterations.
        
        \paragraph{Prompt Construction} We employ a structured template prompt comprising: (1) task statement (i.e., generating floating-point values for symbolic variables to maximize floating-point errors), (2) source function definition, (3) list of symbolic and concrete variables with concrete values applied, (4) current best result, including inputs and measured relative error, and (5) numerical constraints and sample output. Detailed prompts are available in the attached artifact.
        If no improvement is observed after a fixed number of trials (i.e., 10), we further split the symbolic variable set, similar to FPGen. Such a progressive reduction helps avoid stagnation in high-dimensional spaces. If such a split is not possible, we restart the process with the updated reference point. This iterative refinement is capped at one hour.  




    \subsection{Numerical Stabilization}
        
        \subsubsection{Numerical Expression Acquisition}
            \label{s5.2:composition}

            To address challenges with existing benchmarks (discussed in Section \ref{s1:intro}),  we construct a new dataset using three composition strategies that mitigate data contamination, enable control over complexity, and improve generalizability.
            These strategies are designed to systematically vary structural complexity and variable interactions. $f$ is a base expression and $g_i$ are inner expressions, both drawn from the baseline dataset~\cite{herbie}, while $m$ denotes the resulting composed expression.
             \begin{itemize}

                \item \textbf{Unary:} All inputs of $f$ are replaced with single-variable expressions over the same variable, yielding a single-variable composition $f(g_1(x), g_2(x), \ldots)$. 

                    $f(x, y) = x + y - 2$ \& $g_1(x) = x^2$ and $g_2(x) = x + 1$ \\
                    $m(x) = f(g_1(x), g_2(x)) = x^2 + x - 1$

                \item \textbf{Full:} Each variable in $f$ is replaced with a corresponding single-variable expression over that variable, e.g., $f(x, y, \ldots) \rightarrow f(g_1(x), g_2(y), \ldots)$.
                    
                    $f(x, y,z) = x + y - z$ \\
                    $g_1(x) = x^2$, $g_2(y) = log(y)$, and $g_3(z) = \sqrt{z}$ \\
                    $m(x,y,z) = f(g_1(x), g_2(y), g_3(z)) = x^2 + log(y) -\sqrt{z}$

                \item \textbf{Mixed:} A random subset of variables in $f$ is replaced, while others remain unchanged, e.g., $f(x, y, \ldots) \rightarrow f(g_1(x), y, \ldots)$.
                    
                    $f(x, y) = x + y - 2$ \& $g_1(x) = x^2$ and $y$ unchanged\\
                    $m(x,y) = f(g_1(x), y) = x^2 + y -2$  
                    
            \end{itemize}

            We span a spectrum from simple to complex expressions, reducing the likelihood of overlap with existing training data of LLMs. Each composed expression is associated with a valid input domain derived from the original preconditions~\cite{herbie}. These constraints are propagated through the expression structure to determine the final input domain, ensuring validity under operations such as logarithms and square roots. For variables without specified domains, we assign the full \texttt{\small float64} range (i.e., approximate $[-1.79\times10^{308}, 1.79\times10^{308}]$). The resulting domains are then finalized by enforcing standard mathematical constraints (e.g., $x > 0$ for $\log{x}$).
            Due to the combinatorial growth in possible compositions, we sample a fixed number of expressions for each group, where a group is defined by variable count and conditionality (conditional vs. non-conditional). We exclude expressions that already achieve 100\% accuracy, as they cannot be further improved, resulting in a final dataset of \totalExpressions expressions.     
 


        \subsubsection{Experiment Design}
            \label{s5.2:approach}
            \begin{figure}[t]
    \centering
    \includegraphics[width=\linewidth]{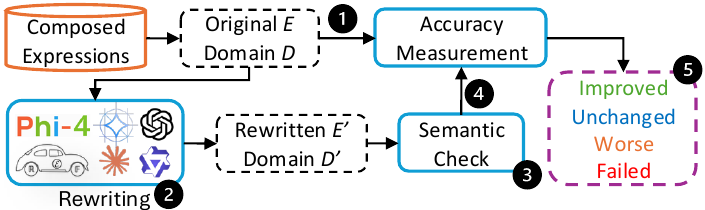}
    \caption{Numerical stabilization workflow}
    \label{fig:Fig2}
    \vspace{-10pt}
\end{figure}    

            Figure \ref{fig:Fig2} illustrates the workflow. For each expression $E$ with domain $D$, we compute its bits of error and accuracy (Step \ding{202}). Herbie~\cite{herbie} is used as the baseline, a tool that produces numerically stable rewrite variants.  Similar to the baseline, we adopt the error metric based on STOKE \cite{STOKE}, which measures error in bits as the $\log_2$ distance between floating-point and high-precision evaluations, with a maximum of 53 bits, corresponding to the significand precision of \texttt{float64} numbers. For a floating-point approximation $x$ and its high-precision counterpart, the error is:


            \begin{center}
                $E(x,y) = \log_2 \lvert \{ z \in \texttt{FP} \mid \min(x,y) \le z \le \max(x,y) \} \rvert$
            \end{center}

            \noindent We evaluate expressions over a minimum of 256 samples per expression and up to 2,560 samples for complex or wide-range domains (10$\times$ more input-space coverage than prior work \cite{herbie}). 
            %
            For unbounded domains, we apply log-uniform sampling to ensure that both very small and very large values are proportionally represented. 
            For ground truth, we use fixed 256-bit precision via \texttt{mpmath} \cite{mpmath}. 
            We also report an accuracy score defined as $100 \times (1 - e/53)$, where $e$ is the number of bits of error, to provide an intuitive and interpretable comparison across expressions~\cite{accuracy}. For instance, an expression $E_1$ with 10.6 bits of error corresponds to 80\% accuracy, while $E_2$ with 53 bits of error yields 0\%, reflecting no matching bits.
            

            In Step \ding{203}, we instruct each model to rewrite the original expression to improve numerical accuracy while preserving semantics. 
            The baseline relies on randomized sampling controlled by an internal seed. 
            We run the baseline five times with different fixed seeds and select the best rewrite among the resulting candidates. Each candidate rewrite then passes through a mathematical semantic-equivalence check (Step \ding{204}, detailed later in \ref{s5.2:semantic}). Only candidates judged equivalent proceed to accuracy evaluation, using the same sampling and error computation (Step \ding{205}), where error is measured as the average bits of error over all sampled inputs.
            



            
            \paragraph{LLM Prompting Technique} For LLMs, we evaluate two prompting strategies: zero-shot and iterative few-shot loop. In both settings, models are provided with the original expression and its variable domain, and are instructed to produce a numerically stable rewrite $E'$ along with updated variable domains $D'$ if the rewrite introduces new mathematical constraints (e.g., non-negative for square root). We follow a structured prompting format consistent with Task 1 and especially enforce preserving the mathematical semantic equivalence during rewrites, and for a few-shot iterative loop, a list of past results, including rewritten expressions, variable domains, error bits, and accuracy. We allow up to five iterations.
            We retain the best-performing result across all few-shot iterations rather than the final output, alongside the five independent zero-shot outputs, for evaluation. Finally, a rewrite is classified as \textit{improved} if it reduces error and increases accuracy for all sampled inputs. 

        \subsubsection{Semantic Equivalence Check}
            \label{s5.2:semantic}

             We verify a LLM-stabilized expression $E'$ of $E$ in two stages.
            \paragraph{Static Structural Analysis} Stage 1 rejects catastrophic semantic changes that include (1) variable drops, which remove a degree of freedom the rewrite cannot recover; and (2) complete removal of all conditional branches of $E$, which collapses piecewise behavior that a flat expression cannot reproduce. Rejected expressions attain a mean accuracy of \avgAccuracyFailedSemantic, versus \avgAccuracyPassedSemantic for those passing, indicating the effectiveness of these checks. Other structural differences, such as newly introduced operations (\texttt{log}, \texttt{exp}), high-precision literals, and altered thresholds or predicates, can still admit an equivalent rewrite (e.g., redundant branches that merge without changing semantics). We therefore postpone the equivalence check for those to stage 2.
            

            \paragraph{Dynamic Numerical Validation} Stage 2 first attempts a symbolic pre-check with SymPy \cite{sympy}. If the difference $E-E'$ reduces to zero within a 5-second timeout, equivalence is proven exactly, and no sampling is needed. Among \totalExpressions evaluated expressions, \percentSTwoSympyOfExprsVTwoPhiOS to \percentSTwoSympyOfExprsVTwoClaudeOpusOS rewrites passed the pre-check.
            This discharges polynomial and rational rewrites cheaply. 
            Note that stabilization is mathematical equivalence, not floating-point semantic equivalence.
            When SymPy cannot decide, whether because the expression is transcendental, fails to parse, or exceeds the timeout, the candidate falls through to numerical testing.  Numerical testing draws points from three sources. A fixed grid of special scalars ($0,\pm1,\pm\pi,\pm\infty$, sub-normal and large magnitudes) covers regions of cancellation, roots, and periodicity that random sampling rarely reaches. Uniform random samples are stratified across five magnitude regimes, i.e., fixed value ranges (near-zero, unit, moderate, large, positive-large), with an equal share drawn from each so that no scale is under-sampled. Control flow is handled separately. For each branch predicate, near-boundary points are sampled at the threshold and at $\pm1$ ULP and $\pm\epsilon$ offsets, since a rewrite that alters a branch condition diverges only within this narrow transition region. 

        \paragraph{Agreement Metric} We quantify disagreement via the gap in units in the last place (ULP),  the number of floating-point values between the two results. We use ULP instead of relative error because relative error becomes unreliable near zero \cite{Goldberg1991, herbie}. Following STOKE \cite{STOKE} and Herbie \cite{herbie}, this count is our error measure for expressions that should be mathematically equal. We treat a gap of up to 4 ULP as rounding and anything larger as a real difference, falling back to a plain $10^{-12}$ cutoff where ULP counting is unreliable. For small nonzero gaps, we recompute the true answer at high precision using a 50-digit \texttt{mpmath}, since a low-precision reference rounds too and can hide the difference~\cite{Kulkarni2025}. This lets us still treat true identities such as $\sin^2 x+\cos^2 x \equiv 1$ as equal, even though they land a few ULP away from 1 in double precision.

        Operations such as division, logarithm, and root are undefined over parts of Real $\mathbb{R}$, so a rewrite can change where an expression is valid \cite{Kleene1952}. At each sampled point, we evaluate the original $E$ and its rewrite $E'$ and check which is undefined. If exactly one does, their domains differ: \emph{expansion} ($E'$ defined where $E$ is not) only adds valid inputs and cannot break callers, so we accept it as a robustness gain; \emph{shrinkage}  drops accepted inputs and is a strict regression. This asymmetry mirrors program refinement, which may accept more inputs but never fewer \cite{Morgan1994}. Stage 2 returns one of four verdicts. \textsc{Equivalent} and \textsc{Not-Equivalent} record agreement or a witnessed disagreement on the shared domain. \textsc{Domain-Shifted} records expansion or shrinkage. \textsc{Error} marks a compile failure due to, for example, syntax issues.  
\section{RESULTS}
\label{s5:result}

We evaluate four LLMs, including open-source, proprietary, and frontier models. 
The local models are deployed via \texttt{Ollama} \cite{ollama} and include the latest versions of Phi4:14B and GPT-OSS:120B. The API-based models include GPT-4o mini (OpenAI) \cite{gpt4omini} and Claude Opus 4.8 with adaptive thinking enabled (Anthropic) \cite{claudeopus4.8}. 
For Task 1, the dataset consisted of 21 source programs, using approximately \taskOnePromptCountFull prompts and roughly \taskOneTokentCount tokens, while for Task 2, the dataset consisted of \totalExpressions expressions, using approximately \taskTwoPromptCountFullNum prompts and roughly \taskTwoTokentCount tokens, for a combined total of approximately \textbf{\totalPromptCount prompts} and over \textbf{\totalTokenCount tokens}.

    \begin{table*}[t]
    \centering
    \caption{Comparison of Relative Error For Model-Generated Inputs Across Benchmarks (Including Results From FPGen \cite{fpgen})}
    \scriptsize
    \setlength{\tabcolsep}{3pt}
    \begin{filecontents*}{Table2_data.csv}
Benchmark,Phi4:14B,GPT4o-mini,GPT-OSS:120B,Claude Opus 4.8,FPGen,Random,S3FP,KLEE-FLOAT
Recursive Sum,\textbf{1.0000e+00},4.6826e-14,2.7246e-06,5.2402e-03,\textbf{1.0000e+00},0.0000e+00,0.0000e+00,0.0000e+00
Pairwise Sum,1.5637e-13,3.2310e-14,\textbf{1.0867e-08},3.9919e-14,1.3174e-16,0.0000e+00,0.0000e+00,0.0000e+00
Compensated Sum,\textbf{1.0000e+00},1.7846e-05,\textbf{1.0000e+00},4.7371e-14,\textbf{1.0000e+00},0.0000e+00,0.0000e+00,0.0000e+00
Vector Sum,1.0000e+00,5.9365e-12,1.0000e+00,\textbf{1.2089e+24},1.0000e+00,0.0000e+00,0.0000e+00,0.0000e+00
Vector 1-Norm,\textbf{3.1433e-16},\textbf{3.1433e-16},\textbf{2.4396e-16},\textbf{2.2180e-16},0.0000e+00,0.0000e+00,0.0000e+00,0.0000e+00
Vector 2-Norm,\textbf{2.6182e-16},\textbf{2.6182e-16},\textbf{2.6182e-16},\textbf{2.2127e-16},\textbf{2.2117e-16},\textbf{3.1216e-16},\textbf{3.1170e-16},0.0000e+00
Vector Dot Product,1.2588e-10,1.2586e-10,1.7582e-06,1.2586e-10,\textbf{1.9190e-04},1.7010e-12,5.5831e-10,0.0000e+00
Vector Convolution,3.8268e-12,1.4961e-09,0.0000e+00,4.3462e-09,\textbf{2.0446e-04},9.2803e-13,1.9864e-10,0.0000e+00
Matrix-Vector Product,3.5806e-10,3.1901e-10,3.9803e-14,4.3820e-14,\textbf{8.9366e-04},0.0000e+00,0.0000e+00,0.0000e+00
Matrix Multiplication,\textbf{3.0126e-04},1.0996e-09,4.3160e-16,1.0996e-09,2.5783e-14,1.1102e-16,1.1102e-16,0.0000e+00
Weighted Mean,1.2291e-13,1.5601e-10,3.7737e-16,1.5601e-10,\textbf{1.0000e+00},9.4290e-12,1.6620e-07,0.0000e+00
Weighted Variance (w),\textbf{2.0708e+00},3.5793e-09,2.9208e-10,3.3107e-09,2.2858e-12,7.9593e-12,2.0918e-05,0.0000e+00
Weighted Variance (m),6.2444e-15,4.2027e-13,0.0000e+00,9.9521e-15,\textbf{7.6280e-02},1.5039e-11,2.5955e-05,0.0000e+00
Weighted Std Dev (w),\textbf{1.0662e+00},3.3764e-10,6.1206e-02,5.1390e-10,1.1429e-12,3.9797e-12,1.0459e-05,0.0000e+00
Weighted Std Dev (m),5.3631e-03,9.1254e-14,1.2276e-14,4.9090e-15,\textbf{3.7439e-02},7.5193e-12,1.2977e-05,0.0000e+00
Weighted Total Sum Sq (m),\textbf{4.6598e-16},\textbf{5.5529e-16},\textbf{3.3350e-16},\textbf{5.9715e-16},\textbf{4.4513e-16},\textbf{5.5294e-16},\textbf{4.7739e-16},0.0000e+00
Weighted Abs Dev (m),1.1529e+18,4.1688e+00,3.0900e-14,\textbf{1.7732e+38},1.0000e+00,2.6840e-11,2.2077e-05,0.0000e+00
Weighted Skewness (m),3.2426e+00,8.0531e-10,1.1271e+00,\textbf{1.5347e+16},2.5675e+01,2.5025e-11,3.1646e-02,0.0000e+00
Weighted Kurtosis (m),\textbf{6.6667e-01},6.1418e-03,3.2518e-15,\textbf{6.6667e-01},1.7733e-12,4.5107e-11,4.3139e-08,0.0000e+00
LU Decomposition,4.9963e-04,5.6105e-06,2.5307e-16,1.2142e-08,\textbf{2.7327e+00},0.0000e+00,0.0000e+00,0.0000e+00
QR Decomposition,\textbf{4.7899e+22},5.3605e-05,\textbf{5.9866e+22},5.2536e-10,2.5912e-14,0.0000e+00,0.0000e+00,0.0000e+00
\end{filecontents*}
\pgfplotstabletypeset[
        col sep=comma,
        string type,
        assign column name/.style={/pgfplots/table/column name={\textbf{#1}}},
        columns/Benchmark/.style={string type,column type=l},
        columns/Phi4:14B/.style={column type=r},
        columns/Gemma3/.style={column type=r},
        columns/GPT4o-mini/.style={column type=r},
        columns/Claude Haiku 4.5/.style={column type=r},
        columns/Claude Opus 4.8/.style={column type=r},
        columns/GPT-OSS:120B/.style={column type=r},
        columns/Qwen3/.style={column type=r},
        columns/FPGen/.style={column type=r},
        columns/Random/.style={column type=r},
        columns/S3FP/.style={column type=r},
        columns/KLEE-FLOAT/.style={column type=r},
        every head row/.style={before row=\toprule,after row=\midrule},
        every last row/.style={after row=\bottomrule}
    ]{Table2_data.csv}
    \label{tab:Table2}
    \vspace{-5pt}
\end{table*}  
    


    \subsection{RQ1: Floating-Point Instability Detection}

        Across benchmarks, LLMs and the baseline each achieve the largest relative errors on nine benchmarks and perform comparably on three others, as shown in Table \ref{tab:Table2}. LLMs also outperform other baseline tools such as random generators, S3FP, and KLEE-FLOAT, but remain inconsistent when compared directly to baseline. This contrast is illustrated by individual cases. In compensated sum, GPT-OSS and Phi4 match the baseline with a relative error of 1.0. In contrast, for weighted variance (m), baseline produces a much larger error (7.63e-02) than all LLMs (4.17e-13), showing clear sensitivity to benchmark structure.

        FPGen explores floating-point error space using symbolic execution and therefore produces errors that are typically 4 to 10 orders of magnitude larger than those produced by LLMs and other tools. In comparison, LLMs are more stochastic. They are effective at generating adversarial inputs and occasionally uncover catastrophic numerical failures that search-based tools miss (Table \ref{tab:Table2}). 
        
        LLMs perform well when benchmarks involve multi-stage statistical formulas or long arithmetic expressions. For example, weighted statistical functions show multiple LLMs' best performance, with Phi4 dominating the benchmarks such as weighted kurtosis, weighted variance (w), and weighted standard deviation (w). Best LLM error magnitudes range from $10^{-1}$ to $10^{0}$, compared to $10^{-12}$ in the baseline. These formulas contain multiple reductions, mean subtraction, exponentiation, and accumulation, which introduce cancellation, overflow, and scaling imbalance. 

        Surprisingly, LLMs consistently underperform on simple linear algebra kernels and reduction operations, such as  matrix-vector product. These benchmarks require high numerical precision, where small cancellation effects can accumulate and significantly affect the final result. The baseline systematically constructs such high-precision-sensitive cases, while LLMs typically remain in the standard precision regime and fail to expose these subtle error amplification behaviors. For example, the best LLM result for vector dot product shows 1.76e-06  in relative error (produced by GPT-OSS), whereas the baseline produces 1.92e-04. 
        
        LLMs can explore rare structural degeneracies that traditional search tools miss. FPGen concretizes a subset of input variables before symbolic execution, exploring a bounded path space. As a result, certain regions can be pruned or under-explored, so they may miss certain higher-error inputs. LLMs complement this by targeting atypical structures, occasionally surfacing higher-error inputs. For example, Phi4 generates inputs that trigger an extremely large relative error (1.15e+18) for the weighted absolute deviation (m), arising from a near-zero denominator in the relative error computation. In QR decomposition, Phi4 and GPT-OSS both discover inputs producing errors around $10^{22}$. These extreme cases are not found by the baseline. Phi4, in particular, identifies multiple extreme cases in benchmarks primarily because its training data include heavily curated, math-centric synthetic data.
        
       
        


        
    \begin{rqbox}
        \textbf{RQ1 Finding.} LLMs are less consistent than FPGen at finding floating-point error-inducing inputs, but can occasionally generate extreme, structurally degenerate cases that classical tools miss, particularly in multi-stage arithmetic and weighted statistical benchmarks.
    \end{rqbox}


    \subsection{RQ2: Semantic Equivalence Checks}

\begin{table*}[t]
    \centering
    \scriptsize
    \setlength{\tabcolsep}{3pt}
    \caption{Summary of Semantic-Equivalence Check in Two Stages for Each Model, Shown for Zero-Shot (0S) and Few-Shot (FS). Stage 1 percentages are computed over the total number of expressions ($N=\totalExprsVTwoClaudeOpusOS$), while Stage 2 percentages are computed over the number of expressions passing Stage 1}
    \label{tab:Table0}
    \begin{tabular}{l r r r r r r r r r r r}
        \toprule
        & \multicolumn{5}{c}{\textbf{Stage 1 --- Static Structural Analysis} (w.r.t. $N$)} & \multicolumn{5}{c}{\textbf{Stage 2 --- Dynamic Numerical Evaluation} (w.r.t. Stage 1 Pass)}\\
        \cmidrule(lr){2-6}\cmidrule(lr){7-11}
        & \multicolumn{4}{c}{\ding{55}\ Stage 1 Reject} & & \multicolumn{2}{c}{\checkmark\ Accepted} & \multicolumn{2}{c}{\ding{55}\ Rejected} & \\
        \cmidrule(lr){2-5}\cmidrule(lr){7-8}\cmidrule(lr){9-10}
        Model & \makecell[r]{Variables\\Removed} & \makecell[r]{Conditions\\Dropped} & Both & \makecell[r]{Other\\Error(s)} & \makecell[r]{Stage 1\\Pass} & Equivalent & \makecell[r]{Domain\\Expanded} & \makecell[r]{Not\\Equivalent} & \makecell[r]{Domain\\Shrink} & \makecell[r]{Stage 2\\Pass}\\
        \midrule
        Claude Opus 0S & \sOneRejVarsVTwoClaudeOpusOS\ (\percentSOneRejVarsVTwoClaudeOpusOS) & \sOneRejCondsVTwoClaudeOpusOS\ (\percentSOneRejCondsVTwoClaudeOpusOS) & \sOneRejBothVTwoClaudeOpusOS\ (\percentSOneRejBothVTwoClaudeOpusOS) & \sOneRejOtherVTwoClaudeOpusOS\ (\percentSOneRejOtherVTwoClaudeOpusOS) & \sOnePassVTwoClaudeOpusOS\ (\percentSOnePassVTwoClaudeOpusOS) & \sTwoEquivVTwoClaudeOpusOS\ (\percentSTwoEquivOfSOnePassVTwoClaudeOpusOS) & \sTwoDomainExpandVTwoClaudeOpusOS\ (\percentSTwoDomainExpandOfSOnePassVTwoClaudeOpusOS) & \sTwoNotEquivVTwoClaudeOpusOS\ (\percentSTwoNotEquivOfSOnePassVTwoClaudeOpusOS) & \sTwoDomainShrinkVTwoClaudeOpusOS\ (\percentSTwoDomainShrinkOfSOnePassVTwoClaudeOpusOS) & \sTwoTotalPassedVTwoClaudeOpusOS\ (\percentSTwoTotalPassedOfSOnePassVTwoClaudeOpusOS)\\
        Claude Opus FS & \sOneRejVarsVTwoClaudeOpusMS\ (\percentSOneRejVarsVTwoClaudeOpusMS) & \sOneRejCondsVTwoClaudeOpusMS\ (\percentSOneRejCondsVTwoClaudeOpusMS) & \sOneRejBothVTwoClaudeOpusMS\ (\percentSOneRejBothVTwoClaudeOpusMS) & \sOneRejOtherVTwoClaudeOpusMS\ (\percentSOneRejOtherVTwoClaudeOpusMS) & \sOnePassVTwoClaudeOpusMS\ (\percentSOnePassVTwoClaudeOpusMS) & \sTwoEquivVTwoClaudeOpusMS\ (\percentSTwoEquivOfSOnePassVTwoClaudeOpusMS) & \sTwoDomainExpandVTwoClaudeOpusMS\ (\percentSTwoDomainExpandOfSOnePassVTwoClaudeOpusMS) & \sTwoNotEquivVTwoClaudeOpusMS\ (\percentSTwoNotEquivOfSOnePassVTwoClaudeOpusMS) & \sTwoDomainShrinkVTwoClaudeOpusMS\ (\percentSTwoDomainShrinkOfSOnePassVTwoClaudeOpusMS) & \sTwoTotalPassedVTwoClaudeOpusMS\ (\percentSTwoTotalPassedOfSOnePassVTwoClaudeOpusMS)\\
        GPT4o-mini 0S & \sOneRejVarsVTwoGPTminiOS\ (\percentSOneRejVarsVTwoGPTminiOS) & \sOneRejCondsVTwoGPTminiOS\ (\percentSOneRejCondsVTwoGPTminiOS) & \sOneRejBothVTwoGPTminiOS\ (\percentSOneRejBothVTwoGPTminiOS) & \sOneRejOtherVTwoGPTminiOS\ (\percentSOneRejOtherVTwoGPTminiOS) & \sOnePassVTwoGPTminiOS\ (\percentSOnePassVTwoGPTminiOS) & \sTwoEquivVTwoGPTminiOS\ (\percentSTwoEquivOfSOnePassVTwoGPTminiOS) & \sTwoDomainExpandVTwoGPTminiOS\ (\percentSTwoDomainExpandOfSOnePassVTwoGPTminiOS) & \sTwoNotEquivVTwoGPTminiOS\ (\percentSTwoNotEquivOfSOnePassVTwoGPTminiOS) & \sTwoDomainShrinkVTwoGPTminiOS\ (\percentSTwoDomainShrinkOfSOnePassVTwoGPTminiOS) & \sTwoTotalPassedVTwoGPTminiOS\ (\percentSTwoTotalPassedOfSOnePassVTwoGPTminiOS)\\
        GPT4o-mini FS & \sOneRejVarsVTwoGPTminiMS\ (\percentSOneRejVarsVTwoGPTminiMS) & \sOneRejCondsVTwoGPTminiMS\ (\percentSOneRejCondsVTwoGPTminiMS) & \sOneRejBothVTwoGPTminiMS\ (\percentSOneRejBothVTwoGPTminiMS) & \sOneRejOtherVTwoGPTminiMS\ (\percentSOneRejOtherVTwoGPTminiMS) & \sOnePassVTwoGPTminiMS\ (\percentSOnePassVTwoGPTminiMS) & \sTwoEquivVTwoGPTminiMS\ (\percentSTwoEquivOfSOnePassVTwoGPTminiMS) & \sTwoDomainExpandVTwoGPTminiMS\ (\percentSTwoDomainExpandOfSOnePassVTwoGPTminiMS) & \sTwoNotEquivVTwoGPTminiMS\ (\percentSTwoNotEquivOfSOnePassVTwoGPTminiMS) & \sTwoDomainShrinkVTwoGPTminiMS\ (\percentSTwoDomainShrinkOfSOnePassVTwoGPTminiMS) & \sTwoTotalPassedVTwoGPTminiMS\ (\percentSTwoTotalPassedOfSOnePassVTwoGPTminiMS)\\
        GPT-OSS 0S & \sOneRejVarsVTwoGPTOSSOS\ (\percentSOneRejVarsVTwoGPTOSSOS) & \sOneRejCondsVTwoGPTOSSOS\ (\percentSOneRejCondsVTwoGPTOSSOS) & \sOneRejBothVTwoGPTOSSOS\ (\percentSOneRejBothVTwoGPTOSSOS) & \sOneRejOtherVTwoGPTOSSOS\ (\percentSOneRejOtherVTwoGPTOSSOS) & \sOnePassVTwoGPTOSSOS\ (\percentSOnePassVTwoGPTOSSOS) & \sTwoEquivVTwoGPTOSSOS\ (\percentSTwoEquivOfSOnePassVTwoGPTOSSOS) & \sTwoDomainExpandVTwoGPTOSSOS\ (\percentSTwoDomainExpandOfSOnePassVTwoGPTOSSOS) & \sTwoNotEquivVTwoGPTOSSOS\ (\percentSTwoNotEquivOfSOnePassVTwoGPTOSSOS) & \sTwoDomainShrinkVTwoGPTOSSOS\ (\percentSTwoDomainShrinkOfSOnePassVTwoGPTOSSOS) & \sTwoTotalPassedVTwoGPTOSSOS\ (\percentSTwoTotalPassedOfSOnePassVTwoGPTOSSOS)\\
        GPT-OSS FS & \sOneRejVarsVTwoGPTOSSMS\ (\percentSOneRejVarsVTwoGPTOSSMS) & \sOneRejCondsVTwoGPTOSSMS\ (\percentSOneRejCondsVTwoGPTOSSMS) & \sOneRejBothVTwoGPTOSSMS\ (\percentSOneRejBothVTwoGPTOSSMS) & \sOneRejOtherVTwoGPTOSSMS\ (\percentSOneRejOtherVTwoGPTOSSMS) & \sOnePassVTwoGPTOSSMS\ (\percentSOnePassVTwoGPTOSSMS) & \sTwoEquivVTwoGPTOSSMS\ (\percentSTwoEquivOfSOnePassVTwoGPTOSSMS) & \sTwoDomainExpandVTwoGPTOSSMS\ (\percentSTwoDomainExpandOfSOnePassVTwoGPTOSSMS) & \sTwoNotEquivVTwoGPTOSSMS\ (\percentSTwoNotEquivOfSOnePassVTwoGPTOSSMS) & \sTwoDomainShrinkVTwoGPTOSSMS\ (\percentSTwoDomainShrinkOfSOnePassVTwoGPTOSSMS) & \sTwoTotalPassedVTwoGPTOSSMS\ (\percentSTwoTotalPassedOfSOnePassVTwoGPTOSSMS)\\
        Phi4 0S & \sOneRejVarsVTwoPhiOS\ (\percentSOneRejVarsVTwoPhiOS) & \sOneRejCondsVTwoPhiOS\ (\percentSOneRejCondsVTwoPhiOS) & \sOneRejBothVTwoPhiOS\ (\percentSOneRejBothVTwoPhiOS) & \sOneRejOtherVTwoPhiOS\ (\percentSOneRejOtherVTwoPhiOS) & \sOnePassVTwoPhiOS\ (\percentSOnePassVTwoPhiOS) & \sTwoEquivVTwoPhiOS\ (\percentSTwoEquivOfSOnePassVTwoPhiOS) & \sTwoDomainExpandVTwoPhiOS\ (\percentSTwoDomainExpandOfSOnePassVTwoPhiOS) & \sTwoNotEquivVTwoPhiOS\ (\percentSTwoNotEquivOfSOnePassVTwoPhiOS) & \sTwoDomainShrinkVTwoPhiOS\ (\percentSTwoDomainShrinkOfSOnePassVTwoPhiOS) & \sTwoTotalPassedVTwoPhiOS\ (\percentSTwoTotalPassedOfSOnePassVTwoPhiOS)\\
        Phi4 FS & \sOneRejVarsVTwoPhiMS\ (\percentSOneRejVarsVTwoPhiMS) & \sOneRejCondsVTwoPhiMS\ (\percentSOneRejCondsVTwoPhiMS) & \sOneRejBothVTwoPhiMS\ (\percentSOneRejBothVTwoPhiMS) & \sOneRejOtherVTwoPhiMS\ (\percentSOneRejOtherVTwoPhiMS) & \sOnePassVTwoPhiMS\ (\percentSOnePassVTwoPhiMS) & \sTwoEquivVTwoPhiMS\ (\percentSTwoEquivOfSOnePassVTwoPhiMS) & \sTwoDomainExpandVTwoPhiMS\ (\percentSTwoDomainExpandOfSOnePassVTwoPhiMS) & \sTwoNotEquivVTwoPhiMS\ (\percentSTwoNotEquivOfSOnePassVTwoPhiMS) & \sTwoDomainShrinkVTwoPhiMS\ (\percentSTwoDomainShrinkOfSOnePassVTwoPhiMS) & \sTwoTotalPassedVTwoPhiMS\ (\percentSTwoTotalPassedOfSOnePassVTwoPhiMS)\\
        \bottomrule
    \end{tabular}
    \vspace{-8pt}
\end{table*}
        Rewrites that survive the 2-stage semantic check are substantially more accurate than those it filters out, achieving a mean accuracy of \avgAccuracyPassedSemantic compared to \avgAccuracyFailedSemantic for rejected candidates only. This reduction in accuracy is indicative of the ability of these checks to exclude rewrites that the LLM produces but diverge from the original expression. Table \ref{tab:Table0} summarizes the semantic-equivalence check results across two stages for each model. The reported counts are semantic checks rather than distinct expressions evaluated: each expression is evaluated over at most five independent zero-shot trials and a five-trial few-shot loop. The total number of checks, therefore, varies by model, for two reasons: (1) the evaluation loop halts early when generated rewrites are duplicated, and (2) some LLM generations are invalid.

        The first stage admits a broadly comparable proportion of candidates across models, ranging from \percentSOnePassVTwoPhiMS to \percentSOnePassVTwoGPTminiMS of total checks, but the composition of its rejections differs substantially. For Claude Opus and GPT-OSS, rejections arise almost exclusively from dropped conditionals (\percentSOneRejCondsVTwoGPTOSSMS-\percentSOneRejCondsVTwoClaudeOpusOS of checked candidates), with variable removal near zero. GPT-4o mini and Phi4 exhibit a different profile, with rejections concentrated in the joint removal of variables and conditions (\percentSOneRejBothVTwoPhiMS-\percentSOneRejBothVTwoGPTminiMS) and syntax errors (\percentSOneRejOtherVTwoPhiMS-\percentSOneRejOtherVTwoGPTminiOS). The contrast reflects a difference in rewriting competence. Claude Opus and GPT-OSS preserve the function's full set of variables and err only by over-flattening its piecewise structure—a recoverable restructuring error—whereas GPT-4o mini and Phi4 discard variables and their associated branches together, removing part of the computation outright. The latter is the more damaging failure, indicating that the smaller models more often fail to retain the expression's essential structure rather than merely reorganizing it incorrectly.
        

        In stage 2, among the structurally sound candidates Stage~1 admits, equivalent stabilized expressions range from \percentSTwoEquivOfSOnePassVTwoPhiOS to \percentSTwoEquivOfSOnePassVTwoClaudeOpusOS across models, with the remainder either computing different values or shifting the valid domain. 
        This spread reflects a difference in the model's ability to produce a structurally faithful rewrite and a numerically faithful one, which are distinct skills that models possess to varying degrees. 
        GPT-4o mini makes the distinction sharpest. It clears Stage~1 more often than any other model (\percentSOnePassVTwoGPTminiOS to \percentSOnePassVTwoGPTminiMS), yet its form-preserving rewrites frequently alter the computed value (around \percentSTwoNotEquivOfSOnePassVTwoGPTminiOS--\percentSTwoNotEquivOfSOnePassVTwoGPTminiMS not equivalent) or narrow the valid domain (\percentSTwoDomainShrinkOfSOnePassVTwoGPTminiOS--\percentSTwoDomainShrinkOfSOnePassVTwoGPTminiMS shrinkage), indicating that it likely reproduces the shape of the original expression rather than accurate computation. 
        The pattern holds across model scale. The larger models (Claude Opus 4.8, GPT-OSS) pass Stage 2 at over \percentSTwoTotalPassedOfSOnePassVTwoClaudeOpusMS of their Stage 1 survivors, against under \percentSTwoTotalPassedOfSOnePassVTwoGPTminiOS for the smaller models, confirming their reasoning superiority.
        

        Domain shifts split cleanly along model scale. The larger models (Claude Opus, GPT-OSS) expand the valid domain far more often than they contract it (expansion \percentSTwoDomainExpandOfSOnePassVTwoClaudeOpusOS--\percentSTwoDomainExpandOfSOnePassVTwoGPTOSSMS versus contraction at most \percentSTwoDomainShrinkOfSOnePassVTwoGPTOSSMS of those passed Stage 1), leaving the admissible input region intact.
        The smaller models show the converse. GPT-4o mini and Phi4 contract the domain in $\sim$ \percentSTwoDomainShrinkOfSOnePassVTwoPhiMS--\percentSTwoDomainShrinkOfSOnePassVTwoGPTminiMS of cases surviving Stage 1. 
        Smaller models, therefore, fail not only by producing incorrect values but by silently restricting the region over which the rewrite remains valid. 
        A larger number excluded in the numerical evaluation phase compared to the structural analysis phase indicates that LLMs generally struggle to reason about the high-precision numerical properties of expressions while trying to preserve their semantics as instructed in the prompt.


    \begin{rqbox}
        \textbf{RQ2 Finding.} Rewrites surviving the full semantic check average \avgAccuracyPassedSemantic accuracy against \avgAccuracyFailedSemantic for rejected ones. Structural and numerical faithfulness prove to be separate model abilities (\percentSOnePassVTwoPhiOS-\percentSOnePassVTwoGPTminiMS of rewrites preserve structure; only \percentSTwoEquivVTwoPhiOS-\percentSTwoEquivVTwoClaudeOpusOS are truly equivalent), indicating a lack of the high-precision numerical reasoning required to present mathematical equivalence. 
    \end{rqbox}    


    \subsection{RQ3: Overall Rewriting Performance}

        

        \begin{figure}
    \centering
    \small
    \begin{tikzpicture}
        \begin{axis}[
            ybar,
            bar width=9pt,
            width=\linewidth,
            height=0.35\linewidth,
            ylabel={\shortstack{Accuracy\\Increase (\%)}},
            ylabel style={align=center},
            symbolic x coords={
                Claude Opus 0S,
                Claude Opus FS,
                GPT4o-mini 0S,
                GPT4o-mini FS,
                GPT-OSS 0S,
                GPT-OSS FS,
                Phi4 0S,
                Phi4 FS,
                Herbie
            },
            xtick=data,
            xticklabel style={
                rotate=30, 
                anchor=east,
                font=\scriptsize
            },
            ymin=0,
            enlarge x limits=0.05,
            nodes near coords,
            nodes near coords align={vertical},
            nodes near coords style={
                font=\scriptsize,
                rotate=45,
                anchor=south west,
                inner sep=1pt,
                text=black,
            },
            every node near coord/.append style={
                /pgf/number format/.cd,
                fixed,
                precision=1,
            },
            enlarge x limits=0.1,
        ]
        
        \addplot table[
            col sep=comma,
            x=Model,
            y=AvgAccuracyIncrease
        ]{Fig4_data.csv};
        
        \end{axis}
    \end{tikzpicture}
    \vspace{-12pt}
    \caption{Average accuracy increase per model vs. baseline on expressions where rewrites improved, shown for zero-shot (0S) and few-shot (FS). Original expressions' average accuracy is \avgAccuracyOriginalImproved}
    \label{fig:Fig4}

    \vspace{-8pt}
\end{figure}  
         
        We assess the numerical stability of the original and rewritten expressions using up to 2560 sampled inputs drawn from the defined variable domains, as in prior work~\cite{herbie}. Figure \ref{fig:Fig4} reports the overall average accuracy improvement across models. Overall, LLMs struggle to keep up with the baselines in stabilizing expressions, both in terms of quantity (number of expressions stabilized) and quality (magnitude of the improvement in accuracy for stabilized expressions). Figure~\ref{fig:Fig5} visualizes these results, showing baseline stabilizing \vennHerbieImprovesPct of the expressions compared to \vennLLMImprovesPct by LLMs, and Figure~\ref{fig:Fig4} shows the average accuracy improvement of \avgAccuracyIncreaseHerbie by Herbie compared to \avgAccuracyIncreaseClaudeOpusMS by the best-performing LLM, i.e., Opus 4.8. Large models such as GPT-OSS and Claude Opus demonstrate comparable performance. Opus 4.8 improves \percentImprovedClaudeOpusOS and \percentImprovedClaudeOpusMS of cases, whereas GPT-OSS also achieves competitive performance, improving \percentImprovedGPTOSSOS and \percentImprovedGPTOSSMS of expressions under zero-shot and few-shot prompting, respectively.

        \begin{wrapfigure}[12]{l}{0.4\columnwidth}
    \centering
    \vspace{-2ex}
    \footnotesize
    \begin{tikzpicture}
        \def\radiusA{1.3}
        \def\radiusB{1.1}
        \def\offset{0.25}
        \coordinate (cA) at (-\offset cm,0);
        \coordinate (cB) at ( \offset cm,0);
        \pgfmathsetmacro{\interX}{((-\offset+\radiusA)+(\offset-\radiusB))/2}
        \begin{scope}
            \clip (cA) circle (\radiusA cm);
            \fill[red!20] (cA) circle (\radiusA cm);
        \end{scope}
        \begin{scope}
            \clip (cB) circle (\radiusB cm);
            \fill[blue!20] (cB) circle (\radiusB cm);
        \end{scope}
        \begin{scope}
            \clip (cA) circle (\radiusA cm);
            \clip (cB) circle (\radiusB cm);
            \fill[red!10!blue!10] (cA) circle (\radiusA);
        \end{scope}
        \draw (cA) circle (\radiusA cm);
        \draw (cB) circle (\radiusB cm);
        \node at ($(cA)+(0,\radiusA cm)+(-1,-2.5)$) {A};
        \node at ($(cB)+(0,\radiusB cm)+(0.7,-2.2)$) {B};
        \node at ($(cA)+(-0.86,0)$) {\vennHerbieOnly};
        \node at ($(cB)+(0.95,0)$) {\vennLLMOnly};
        \node[align=center] at (\interX,0) {C\\[\baselineskip]\vennBoth};
        \def\boxleft{-\offset-\radiusA-0.2}
        \def\boxright{\offset+\radiusB+0.3}
        \def\boxtop{\radiusA+0.3}
        \def\boxbot{-\radiusA-0.2}
        \draw[rounded corners] (\boxleft,\boxbot) rectangle (\boxright,\boxtop);
        \node[anchor=north east] at (\boxright-0.02,\boxtop-0.04) {Total: \vennTotal};
    \end{tikzpicture}%
    \vspace{-2ex}
    \caption{Expressions with improved accuracy under Herbie (A) and the LLMs (B); the intersection (C) denotes expressions improved by both}
    \label{fig:Fig5}
\end{wrapfigure}

        

        Among smaller models, Phi4 outperforms GPT-4o mini in stabilizing expressions, achieving an average accuracy improvement of \avgAccuracyIncreasePhiOS-\avgAccuracyIncreasePhiMS compared to GPT-4o mini's \avgAccuracyIncreaseGPTminiOS–\avgAccuracyIncreaseGPTminiMS across both prompting techniques. Phi4's stronger performance on the standard MATH benchmark relative to GPT-4o mini \cite{phi4report} helps explain this difference.
        We also observe that the few-shot loop enhances numerical reasoning robustness compared to zero-shot prompting, as conditioning on multiple past input-output iterations provides additional context for improved reasoning and more accurate rewrites. 
        For example, Phi4 reduces its failure rate from \percentFailedPhiOS to \percentFailedPhiMS. 

        \textit{LLMs Outperforming Baselines.} Out of \totalExpressions expressions, \totalLLMBetterThanHerbie (\percentLLMBetterThanHerbie) are stabilized by both the LLM and the baseline, but the LLM achieves higher accuracy. These are cases where the baseline finds \emph{a} valid improvement but not the \emph{best} one. The gap widens with expression size. As the number of variables grows, the baseline's term-local search must improve each region separately and tends to settle for a partial fix, whereas the LLM pattern-matches a transformation that applies to the whole expression and removes the error source in one go. 
        
        Baseline fails to improve accuracy on \totalHerbieNotImproved expressions (\percentHerbieNotImproved). Notably, among these baseline-failed cases, \totalLLMImprovedWhenHerbieNot expressions (\percentLLMImprovedWhenHerbieNot of baseline failures, \percentLLMImprovedWhenHerbieNotOverall overall) are nevertheless improved by at least one LLM.  These expressions share two properties. First, the stabilizing rewrite is a single named identity that spans several library functions, for example, hyperbolic substitution $(e^x-e^{-x})/(e^x+e^{-x}) \rightarrow \tanh(x)$ or logarithmic reformulation $\log(-1+1/z) \rightarrow \log((1-z)/z)$. Second, they exhibit catastrophic cancellation that only such a function-level identity can remove. The baseline composes small rewrites over arithmetic primitives, and its rule set contains no identities spanning functions like $\mathrm{atanh}$ or $\cos 2x$, so no sequence of its local steps reaches the stabilizing form, and it returns the original or a worse variant. LLMs, in contrast, pattern-match these identities from their training data and apply them directly.

     \begin{rqbox}
        \textbf{RQ3 Finding.} Overall, baseline outperforms LLMs on both quality and quantity. LLMs offer complementary value in stabilizing \percentLLMImprovedWhenHerbieNot of the cases where the baselines fail, offering additional advantages due to their symbolic reasoning and atypical structural rewrites. 
    \end{rqbox}


    \begin{figure}[t]
\centering
\resizebox{\columnwidth}{!}{%
\begin{tikzpicture}
\begin{axis}[
    width=\columnwidth,
    height=0.4\columnwidth,
    scale only axis,
    colormap={blues}{rgb255=(230,241,251) rgb255=(133,183,235) rgb255=(55,138,221) rgb255=(24,95,165) rgb255=(4,44,83)},
    colorbar,
    colorbar style={ylabel={Accuracy (\%)}, width=0.22cm, ylabel style={font=\large}, ytick={0,10,20,30}, yticklabel style={font=\normalsize}},
    point meta min=0,
    point meta max=30,
    xtick={0,1,2,3,4,5,6,7,8,9,10},
    xticklabels={1,2,3,4,5,6,7,8,9,10,16},
    xticklabel style={font=\normalsize},
    ytick={0,1,2,3,4,5,6,7,8},
    yticklabels={
        Claude Opus 0S, Claude Opus FS,
        GPT4o-mini 0S, GPT4o-mini FS,
        GPT-OSS 0S, GPT-OSS FS,
        Herbie,
        Phi4 0S, Phi4 FS
    },
    yticklabel style={font=\normalsize},
    y dir=reverse,
    xlabel={Variable Count},
    xlabel style={font=\large},
    enlargelimits=false,
    axis on top,
    xtick align=outside,
    ytick align=outside,
    tick style={line width=0.3pt},
]
\addplot[
    matrix plot*,
    point meta=explicit,
    mesh/cols=11,
] table[x=x, y=y, meta=val, col sep=comma] {heatmap_data.csv};
\end{axis}
\end{tikzpicture}%

}
\vspace{-20pt}
\caption{Average accuracy improvement per model per variable count. Zero-shot (0S) and few-shot (FS) variants are shown side by side for each base model. Accuracy computation is explained in Section \ref{s5.2:approach}}
\label{fig:Fig6}

\vspace{-10pt}
\end{figure}

    \subsection{RQ4: Structure Complexity}

        


         
        \subsubsection{Variable Count}
            
            

            Our initial hypothesis is that LLM performance degrades as the number of variables increases. However, the results do not exhibit a consistent monotonic trend. Figure \ref{fig:Fig6} presents the distribution of accuracy improvement across variable counts for each model. The LLMs achieve their strongest gains on 4-variable expressions (up to \avgAccuracyIncreaseFourVarGPTOSSMS for GPT-OSS) and on expressions with 7 to 10 variables, with weaker improvement in between. This non-monotonic pattern is consistent with the fact that expression complexity also depends on the number of operations and conditional branches rather than variables alone.
            In contrast, baseline is notably stable, attaining comparable gains across all variable counts, ranging from \avgAccuracyIncreaseSixteenVarHerbie to \avgAccuracyIncreaseSevenVarHerbie.
           
            The LLMs' behavior is most erratic for the smaller models. Phi4 and GPT-4o mini produce gains that are scattered unevenly across variable counts rather than concentrated in a consistent range. 
            Larger models behave more predictably. Claude Opus and GPT-OSS offer stronger gains on higher variable counts, but weaker gains on the expressions with fewer variables. 
            Larger expressions offer more structural redundancy and more opportunities to apply known identities, which higher-capacity models recognize more reliably, whereas on single-variable expressions, there is little room to improve over an already-compact form.

            


        \subsubsection{Arithmetic Operations}
        
            \setlength{\tabcolsep}{2pt}
\begin{table}
    \centering
    \caption{Operation Count Statistics in Original Expressions by Variable Count}
    \scriptsize
    
    \begin{tabularx}{\columnwidth}{>{\raggedright\arraybackslash}X *{11}{>{\raggedleft\arraybackslash}X}}        
        \toprule 
        & \multicolumn{11}{c}{\textbf{Variable Count}}\\  
        \cmidrule{2-12}

        & 1 & 2 & 3 & 4 & 5 & 6 & 7 & 8 & 9 & 10 & 16\\

        \midrule
        
        \textbf{Min} & \minOpsOneVar & \minOpsTwoVar & \minOpsThreeVar & \minOpsFourVar & \minOpsFiveVar & \minOpsSixVar & \minOpsSevenVar & \minOpsEightVar & \minOpsNineVar & \minOpsTenVar & \minOpsSixteenVar\\

        \textbf{Max} & \maxOpsOneVar & \maxOpsTwoVar & \maxOpsThreeVar & \maxOpsFourVar & \maxOpsFiveVar & \maxOpsSixVar & \maxOpsSevenVar & \maxOpsEightVar & \maxOpsNineVar & \maxOpsTenVar & \maxOpsSixteenVar\\
        
        \textbf{Avg} & \avgOpsOneVar & \avgOpsTwoVar & \avgOpsThreeVar & \avgOpsFourVar & \avgOpsFiveVar & \avgOpsSixVar & \avgOpsSevenVar & \avgOpsEightVar & \avgOpsNineVar & \avgOpsTenVar & \avgOpsSixteenVar\\


        \bottomrule 
    \end{tabularx}
    \label{tab:Table4}
    \vspace{-10pt}
\end{table}

            Expressions with fewer variables can be more complex due to a greater number of operations, which may reduce the effectiveness of rewriting. Across our dataset, expressions contain an average of \avgOpsOverall operations, with a maximum of \maxOpsOverall.      This observation helps explain the non-monotonic trends reported in the previous subsection. Table \ref{tab:Table4} shows the arithmetic operation count statistics in original expressions by variable count. For example, 6-variable expressions contain an average of \avgOpsSixVar operations, and go up to \maxOpsSixVar, compared to \avgOpsTenVar for 10-variable expressions with a maximum of \maxOpsTenVar. 
            This difference is reflected in performance, where LLM performance with 10-variable expressions achieves higher accuracy improvement (\avgAccuracyIncreaseTenVarLLM) than that with 6-variable expressions (\avgAccuracyIncreaseSixVarLLM).  





        \subsubsection{Composition Strategy}
            LLMs achieve accuracy improvements of \avgAccuracyIncreaseUnaryLLM, \avgAccuracyIncreaseFullLLM, and \avgAccuracyIncreaseMixedLLM for unary, full, and mixed composition strategies, respectively, compared to the baseline improvements of \avgAccuracyIncreaseUnaryHerbie, \avgAccuracyIncreaseFullHerbie, and \avgAccuracyIncreaseMixedHerbie.
            Interestingly, LLMs on full composition (most complex) yield greater improvement than on mixed. Although full-composed expressions contain more operations, all variables are transformed, and their domains are jointly constrained through composition. This often results in tighter effective domains and enables more stable rewrites. Furthermore, the composition process often introduces algebraic simplifications, such as term cancellation, yielding more structured expressions.

    \subsubsection{Control Flow}

        

        
        \begin{figure}
    \centering
    \begin{tikzpicture}
    \begin{axis}[
        ybar,
        bar width=5pt,
        width=\columnwidth,
        height=0.35\columnwidth,
        ylabel={\shortstack{Accuracy\\Increase (\%)}},
        ylabel style={align=center, font=\small},
        symbolic x coords={
            Claude Opus 0S,
            Claude Opus FS,
            GPT4o-mini 0S,
            GPT4o-mini FS,
            GPT-OSS 0S,
            GPT-OSS FS,
            Phi4 0S,
            Phi4 FS,
            Herbie,
        },
        xtick=data,
        xticklabel style={rotate=30, anchor=east, font=\scriptsize},
        ymin=0,
        enlarge x limits=0.07,
        legend style={
            at={(0.5,1.5)},
            anchor=north,
            legend columns=2,
            font=\small
        },
        legend cell align=left,
        nodes near coords,
        nodes near coords style={font=\scriptsize, rotate=90, anchor=west, text=black},
        every node near coord/.append style={/pgf/number format/fixed, /pgf/number format/precision=1},
    ]
    \addplot table[col sep=comma, x=Model, y=Conditional]{Fig7_data.csv};
    \addplot table[col sep=comma, x=Model, y=Unconditional]{Fig7_data.csv};
    \legend{Conditional, Unconditional}
    \end{axis}
\end{tikzpicture}
\vspace{-22pt}
    \caption{Average accuracy increase per model vs. baseline for conditional vs. non-conditional expressions. Zero-shot (0S) and Few-shot (FS). Average accuracy of the  conditional expressions is \avgAccuracyOriginalCond, and non-conditional is \avgAccuracyOriginalNonCond}
    \label{fig:Fig7}

    \vspace{-5pt}
\end{figure}

        In Section \ref{s3:motivation}, Case 2 demonstrates that nested conditional expressions degrade LLM performance due to multiple layers of branching, which challenge the models’ pattern-matching capabilities. Our results show that conditional structure can impact LLM rewriting patterns and performance. Our dataset presents expressions with up to \maxConditionalLevel levels of nesting. LLMs achieve higher accuracy on non-conditional expressions, as nested or multi-level conditionals increase the difficulty of pattern recognition and numerical reasoning. LLMs improve the accuracy of conditional expressions by \avgAccuracyIncreaseCondLLM (from \avgAccuracyOriginalCond), compared to non-conditional cases with accuracy increasing by \avgAccuracyIncreaseUncondLLM (from \avgAccuracyOriginalNonCond). Compared to baseline, with accuracy improvement of \avgAccuracyIncreaseCondHerbie and \avgAccuracyIncreaseNonCondHerbie,
         LLM performance is lower in both cases. Figure \ref{fig:Fig7} compares improvements across models and prompting strategies.

    \subsubsection{High-Precision Literals}

\begin{table}
    \centering
    \caption{Model-wise percentage of expressions with improved accuracy after rewriting for high-precision literal cases}

    \begin{tabularx}{\columnwidth}{l *{2}{>{\centering\arraybackslash}X}}
        \toprule 

       \textbf{Models} & Zero-shot & Few-shot \\

        \midrule
        
        Claude Opus 4.8 & \percentImprovedHighPrecisionClaudeOpusOS & \percentImprovedHighPrecisionClaudeOpusMS \\

        
        GPT4o-mini & \percentImprovedHighPrecisionGPTminiOS &  \percentImprovedHighPrecisionGPTminiMS \\
        
        GPT-OSS:120B & \percentImprovedHighPrecisionGPTOSSOS & \percentImprovedHighPrecisionGPTOSSMS \\
        
        Phi4:14B & \percentImprovedHighPrecisionPhiOS & \percentImprovedHighPrecisionPhiMS \\
        
        
        Herbie & -- & \percentImprovedHighPrecisionHerbie\\
            
        \bottomrule 
    \end{tabularx}
    \label{tab:Table5}
    \vspace{-10pt}
\end{table}


        Scientific software frequently involves high-precision numerical constants. 
        Across the dataset, \totalHighPrecisionOriginal expressions (\percentHighPrecisionOriginal) contain high-precision literals. Table \ref{tab:Table5} summarizes the proportion of cases where rewriting improves accuracy. LLMs stabilize a smaller portion of such expressions, ranging from \percentImprovedHighPrecisionPhiMS to \percentImprovedHighPrecisionGPTOSSOS, compared to \percentImprovedHighPrecisionHerbie for the baseline.  This gap is explained by how high-precision literals are tokenized and represented in LLMs. For example, a constant such as {\tt-5.99633501014107895267e1} is fragmented into a sequence of 12 tokens {\tt ['-', '5', '.', '996', '335', '010', '141', '078', '952', '67', 'e', '1']} by GPT-OSS:120B, determined by which digit sequences were frequent during training rather than by mathematical structure. The semantic role of {\tt e-1} as a $10^{-1}$ multiplier is not preserved, and the model must reconstruct magnitude from co-occurrence statistics, which fails past a few significant digits. Reproducing such a constant faithfully through a rewrite is therefore error-prone, and any deviation in even one digit changes the value of the expression and eliminates the accuracy gain that a correct structural rewrite would otherwise provide. The baseline treats these literals as exact numerical values and preserves them unchanged, which is why it achieves a significantly higher improvement rate.

    \begin{rqbox}
        \textbf{RQ4 Finding.} 
        LLMs perform better on multi-variable expressions with fewer arithmetic operations, where pre-trained algebraic identities apply, and on tighter domains derived via composition strategies, but degrade on cases with control flow and high-precision literals.
    \end{rqbox}
\section{DISCUSSION}
\label{s6:discussion}

        \textbf{Impact of Quantized Models on Numerical Precision.}
        Models in the Ollama family are quantized, typically operating with 4–8 bit precision, whereas API-based models such as GPT-4o mini and Claude do not disclose their internal precision. Quantized models represent weights and activations with reduced bit width (e.g., 16-bit or 8-bit) to improve training and inference efficiency. This reduction in numerical precision, however, can constrain the model’s ability to perform computations that require high accuracy, potentially decreasing reliability when evaluating or transforming floating-point expressions, and it is an interesting direction for future work.  The effect is particularly pronounced in tasks where small variations in intermediate values can accumulate, significantly influencing the final results.

        \textbf{Hybrid Approach.}
            Our results suggest that LLMs and existing floating-point tools offer complementary strengths, motivating a hybrid approach in which LLMs propose candidate inputs or rewrites, and tools such as Herbie or FPGen validate, refine, or further optimize these candidates. For example, LLM-generated rewrites can be passed to Herbie for accuracy improvement, while FPGen can be used to identify adversarial inputs that stress-test LLM outputs. Such a pipeline combines the structural exploration capability of LLMs with the numerical rigor of existing tools. 

            \textbf{Discrepancy Between Intermediate and Final Errors.}
            Floating-point errors may arise in intermediate computations and propagate through subsequent operations, as also observed in FPGen \cite{fpgen}. However, final relative error alone does not fully capture this behavior, since errors introduced earlier can be partially or completely canceled (e.g., when adding two values of similar magnitude but opposite signs). To account for this, in Task 1, we measure not only the final relative error but also the maximum relative error observed at any intermediate step. Our results show that, for many benchmarks, intermediate errors can be significantly larger than the final reported error. For instance, in the recursive summation example discussed earlier, although the final relative error for the Phi4-generated input is $1.4822\times10^{-13}$, the maximum intermediate relative error reaches as high as $1.0$ during execution. This discrepancy indicates that relying solely on final outputs can underestimate the severity of numerical instability.
        
     \textbf{Threats to Validity.}
            While our study provides insights into LLM performance on numerical tasks, several limitations could affect generalizability. First, we evaluated only a subset of the complete dataset since the full composition dataset is exponentially large. This subset was selected to cover diverse and challenging cases, providing representative insights while keeping the evaluation tractable. Second, we focused on four widely used and best-performing LLMs, leaving many other models unexplored. Given the rapid evolution and continuous deployment of new models, future studies could extend evaluation to additional LLMs. Third, our analysis relied on two baseline methods for comparison. Other numerical baselines could influence observed performance differences. Finally, we validate the equivalence of LLM-stabilized expressions using empirical measurements on carefully sampled inputs, which may not be complete. Note that the mathematical equivalence of expressions with transcendental functions is undecidable.

\section{BACKGROUND \& RELATED WORK}
\label{s2:related_work}




        The IEEE-754 \cite{IEEE754, Goldberg1991} formally defines the representation and semantics of floating-point, including rounding modes, overflow behavior, and special values. These works form the theoretical basis for understanding numerical error propagation and the limitations of finite-precision computation.
        


    \textbf{\em Floating-Point Analysis and Stabilization.}
        A large body of work has focused on automatically analyzing and improving numerical accuracy in floating-point programs. Herbie \cite{herbie} comprises a series of work on detecting and stabilizing inaccurate expressions. It synthesizes improved expressions that reduce numerical error by searching for algebraically equivalent rewrites. Daisy \cite{daisy} provides static analysis techniques to compute sound error bounds for floating-point programs. FPGen \cite{fpgen} generates floating-point expressions to stress-test numerical analysis tools. FPTaylor \cite{fptaylor} uses symbolic Taylor expansions and optimization techniques to derive tight error bounds. Herbgrind \cite{herbgrind} complements static methods by dynamically tracing floating-point executions to identify sources of numerical instability. These tools emphasize analysis, optimization, and debugging of floating-point errors, often focusing on either static guarantees or dynamic error attribution. While these tools represent decades of static, dynamic, rule- and heuristic-based development, LLMs have not yet been explored as alternatives to such approaches. This motivates our study of LLMs as a complementary approach for detecting and mitigating floating-point errors.
        

        
        



    \textbf{\em Floating-Point Benchmarking.}
        Benchmark suites have been developed to evaluate the numerical correctness and performance of floating-point systems. FPBench \cite{fpbench} defines a standardized representation for floating-point benchmarks to support reproducible evaluation of accuracy and optimization techniques. Wang et al. \cite{Wang2007} introduce benchmark suites for decimal floating-point applications spanning financial and commercial workloads such as banking, risk management, and billing systems. Laguna et al. \cite{Laguna2022} present HPC-oriented benchmark suites covering proxy applications across MPI, OpenMP, and performance-portability frameworks. These benchmarks primarily focus on system evaluation, numerical correctness, and performance across representative workloads, rather than program transformation or rewriting quality.
        



    \textbf{\em LLMs for Mathematics and Symbolic Reasoning.}
        Recent advances in LLMs have demonstrated strong capabilities in mathematical reasoning. 
        Program-of-Thoughts \cite{Chen2023} further separates reasoning from computation by generating executable code for numerical evaluation. Minerva \cite{Lewkowycz2022} shows that large-scale training on technical corpora enables LLMs to solve challenging mathematical and scientific problems.
        Toolformer \cite{Schick2023} enables models to autonomously invoke external tools such as calculators to improve reasoning reliability. These approaches primarily target the correctness of final answers rather than numerical stability or floating-point representation issues. 






    \textbf{\em LLMs for Scientific Computing.}
        LLMs have also been applied to broader scientific and quantitative reasoning tasks. 
        PAL and accompanied tool-augmented reasoning approaches \cite{Gao2023} improve accuracy by executing generated programs during inference. NumericBench \cite{NumericBench} focuses specifically on numerical capabilities such as arithmetic, comparison, and multi-step reasoning. LLM-SRBench \cite{LLM-SRBench} evaluates scientific equation discovery tasks across multiple domains. These benchmarks mainly assess general numeric reasoning—whether models can correctly perform computations or manipulate symbolic expressions—without explicitly evaluating floating-point stability under finite-precision arithmetic.

\section{CONCLUSION}
\label{s8:conclusion}

    Large language models are increasingly embedded in real-world systems across domains. However, their ability to detect numerical instability and produce stable reformulations remains largely unexplored. We study whether LLMs can approximate systematic numerical analysis and stabilize numerical expressions. We find that LLMs outperform baseline methods precisely where the latter fail and offer greater stability than the baseline across \percentLLMBetterThanHerbie(\totalLLMBetterThanHerbie) of all expressions. Still, LLMs struggle with control flow and high-precision literals, consistently removing such structures rather than reasoning about their numerical implications. Overall, these findings suggest that LLMs offer a meaningful complementary advantage when combined with classical numerical stabilization methods. In particular, LLMs can reduce the reliance on manually derived rules and heuristics that classical tools require, while classical methods can compensate for LLMs' limitations in high-precision and control flow reasoning.
    
    \noindent\textbf{Data Availability.} We have made our code and dataset publicly available at \url{https://anonymous.4open.science/r/LLMNumberICSE26/}.

\bibliographystyle{IEEEtran}
\bibliography{bib} 

\end{document}